\pdfoutput=1
\newcommand*{\ATLASLATEXPATH}{latex/}
\documentclass[UKenglish,cernpreprint,txfonts,atlasstyle,texlive=2011]{\ATLASLATEXPATH atlasdoc}

\usepackage[subfigure]{\ATLASLATEXPATH atlaspackage}
\usepackage{amsmath}
\usepackage{float}
\usepackage{rotating}
\usepackage{\ATLASLATEXPATH atlasbiblatex}

\usepackage{\ATLASLATEXPATH atlascontribute}

\usepackage{\ATLASLATEXPATH atlasphysics}

\graphicspath{{logos/}{figures/}}

\AtlasTitle{Measurements of $\psi(2S)$ and $X(3872)\rightarrow J/\psi\pi^+\pi^-$ production in $pp$ collisions at $\sqrt{s}$ = 8 TeV with the ATLAS detector}

\author{The ATLAS Collaboration}

\AtlasRefCode{BPHY-2015-03}

 \PreprintIdNumber{CERN-EP-2016-193}

 \AtlasJournal{JHEP}

\AtlasAbstract{Differential cross sections are presented for the prompt and non-prompt production of the hidden-charm states $X(3872)$ and  $\psi(2S)$, in the decay mode $J/\psi \pi^+\pi^-$, measured using 11.4~fb$^{-1}$ of $pp$ collisions at $\sqrt{s} = 8$~TeV by the ATLAS detector at the LHC. The ratio of cross-sections $X(3872)/\psi(2S)$ is also given, separately for prompt and non-prompt components, as well as the non-prompt fractions of $X(3872)$ and $\psi(2S)$. Assuming independent single effective lifetimes for non-prompt $X(3872)$ and $\psi(2S)$ production gives $R_B =  \frac{\mathcal{B}(B \rightarrow X(3872)\textrm{ + any}) \mathcal{B}(X(3872) \rightarrow \jpsi\pi^+\pi^-)}{\mathcal{B}(B \rightarrow \psi(2S)\textrm{ + any}) \mathcal{B}(\psi(2S) \rightarrow \jpsi\pi^+\pi^-)} = (3.95 \pm 0.32 \mathrm{(stat)} \pm 0.08\mathrm{(sys)})  \times 10^{-2}$, while separating short- and long-lived contributions, assuming that the short-lived component is due to $B_c$ decays, gives $R_B = (3.57 \pm 0.33\mathrm{(stat)} \pm 0.11\mathrm{(sys)}) \times 10^{-2}$, with the fraction of non-prompt $X(3872)$ produced via $B_c$ decays for $p_\mathrm{T}(X(3872)) > 10$~GeV being $(25 \pm 13\mathrm{(stat)} \pm 2\mathrm{(sys)} \pm 5\mathrm{(spin)})\%$. The distributions of the dipion invariant mass in the $X(3872)$ and  $\psi(2S)$ decays are also measured and compared to theoretical predictions.
}

\hypersetup{pdftitle={ATLAS document},pdfauthor={The ATLAS Collaboration}}

\begin{document}

\maketitle


\section{Introduction}
\label{sec:intro}

The hidden-charm state $X(3872)$ was discovered by the Belle Collaboration in 2003~\cite{belleDisc} through its decay to $J/\psi \pi^{+}\pi^{-}$ in the exclusive decay $B^{\pm} \rightarrow K^{\pm}J/\psi\pi^+\pi^-$. Its existence was subsequently confirmed by CDF~\cite{cdfDisc} through its production in $p\bar{p}$ collisions, and its production was also observed by the BaBar~\cite{babarDisc} and D0~\cite{d0Disc} experiments shortly after. CDF determined~\cite{Abulencia:2006ma} that the only possible quantum numbers for $X(3872)$ were $J^{PC}=1^{++}$ and $2^{-+}$. At the LHC, the $X(3872)$ was first observed by the LHCb Collaboration~\cite{lhcbObs}, which finally confirmed its quantum numbers to be $1^{++}$~\cite{lhcbQN}. A particularly interesting aspect of the $X(3872)$ is the closeness of its mass, $3871.69 \pm 0.17 \MeV$~\cite{PDGBcTau}, to the $D^{0}\bar{D}^{*0} $ threshold, such that it was hypothesised to be a $D^{0}\bar{D}^{*0}$ molecule with a very small binding energy~\cite{binding}.  A cross-section measurement of promptly produced $X(3872)$ was performed by CMS~\cite{cmsMeas} as a function of $\pt$, and showed the non-relativistic QCD (NRQCD) prediction~\cite{Artoisenet:2009wk} for prompt $X(3872)$ production, assuming a $D^{0}\bar{D}^{*0} $ molecule, to be too high, although
the shape of the $\pt$ dependence was described fairly well.
A later interpretation of $X(3872)$ as a mixed  $\chi_{c1}(2P) \textrm{--}D^{0}\bar{D}^{*0}$ state, where the $X(3872)$ is produced predominantly through its
$\chi_{c1}(2P)$ component, was adopted in conjunction with the next-to-leading-order (NLO) NRQCD model and fitted to CMS
data, showing good agreement~\cite{chidd}.

ATLAS previously observed the $X(3872)$ state while measuring the cross section of prompt and non-prompt $\psi(2S)$ meson
production in the $J/\psi\pi^+\pi^-$ decay channel with 2011 data at a centre-of-mass energy
$\sqrt{s}= 7 \TeV$~\cite{BPHY-2013-06}.
ATLAS later performed cross-section measurements for $J/\psi$ and
$\psi(2S)$ decaying through the $\mu^+\mu^-$ channel at $\sqrt{s} = 7 \TeV$
and $\sqrt{s} = 8 \TeV$~\cite{BPHY-2012-02b}.

In this analysis, a measurement of the differential cross sections for the
production of $\psi(2S)$ and $X(3872)$ states in the decay
channel $J/\psi\pi^+\pi^-$ is performed, using $11.4$ fb$^{-1}$ of proton--proton
collision data collected by the ATLAS experiment at the LHC at $\sqrt{s} = 8 \TeV$.
The  $J/\psi\pi^+\pi^-$ final state allows good invariant mass resolution
through the use of a constrained fit, and provides a straightforward way of comparing the production characteristics of $\psi(2S)$ and $X(3872)$ states, which are fairly close in mass.
The prompt and non-prompt contributions for $\psi(2S)$ and $X(3872)$ are separated, based on an analysis of
the displacement of the production vertex.
Non-prompt production fractions for $\psi(2S)$ and $X(3872)$ are measured, and the $X(3872)/\psi(2S)$ production ratios are measured separately for prompt and non-prompt components.
The non-prompt results show that while the non-prompt $\psi(2S)$ data is readily described by a traditional single-effective-lifetime fit, there are indications in the non-prompt $X(3872)$ data which suggest introducing a two-lifetime fit with both a short-lived and long-lived component. Results are presented here based on both the single- and two-lifetime fit models. In the two-lifetime case, assuming that the short-lived non-prompt component of $X(3872)$ originates from the decays of $B_c$ mesons, the best-fit fractional contribution of the $B_c$ component is determined. The distributions of the dipion invariant mass in $\psi(2S) \to \jpsi\pi^+\pi^-$ and $X(3872) \to \jpsi\pi^+\pi^-$ decays are also measured. Comparisons are made with theoretical models and available experimental data.

\section{The ATLAS detector}
\label{sec:detector}

The ATLAS detector~\cite{PERF-2007-01} is a cylindrical, forward-backward
symmetric, general-purpose particle detector. The innermost part of the inner
detector (ID) comprises pixel and silicon microstrip (SCT) tracking technology
for high-precision measurements, complemented further outwards by the
transition radiation tracker (TRT).
The inner detector spans the pseudorapidity{\footnote{ATLAS uses a right-handed coordinate system with its origin at the nominal interaction point (IP) in the centre of the detector and the $z$-axis along the beam pipe. The $x$-axis points from the IP to the centre of the LHC ring, and the $y$-axis points upward. Polar coordinates $(r,\phi)$ are used in the transverse plane, $\phi$ being the azimuthal angle around the $z$-axis. The pseudorapidity $\eta$ is defined in terms of the polar angle $\theta$ as $\eta = -\ln \tan (\theta/2)$, and the transverse momentum $\pt$ is defined as $\pt = p\sin \theta$. The rapidity $y$ is defined as $y=0.5 \ln [(E+p_z)/(E-p_z)]$, where $E$ and $p_z=p\cos\theta$ refer to energy and longitudinal momentum, respectively.}} range $|\eta|<2.5$ and is immersed in a 2~T
axial magnetic field. Enclosing the ID and the solenoidal magnet are the
electromagnetic and hadronic sampling calorimeters, which
provide good containment of the electromagnetic and hadronic showers in order
to limit punch-through into the muon spectrometer (MS).
Surrounding the calorimeters, the MS covers the rapidity range
$|\eta| < 2.7$ and utilises three air-core toroidal
magnets, each consisting of eight coils, generating a magnetic field providing 1.5--7.5~T$\cdot$m of bending power.
The MS consists of fast-trigger detectors (thin-gap chambers and resistive
plate chambers) as well as precision-measurement detectors (monitored drift
tubes and cathode strip chambers).

The ATLAS detector uses a three-level trigger system in order to select 300~Hz of interesting events to be written out from the 20~MHz of proton bunch collisions. This analysis uses a dimuon trigger with the lowest available transverse momentum threshold of $4$~GeV for each muon. The level-1 muon trigger finds regions-of-interest (RoIs) by searching for hit coincidences in layers of the muon trigger detectors inside predefined geometrical windows. The software-based two-stage high-level trigger (HLT) is seeded by the level-1 RoIs, and uses more precise MS and ID information to reconstruct the final muon trigger objects with a resolution comparable to the full offline reconstruction.

\section{Event selection}
\label{sec:eventsel}

Events used in this analysis are triggered by a pair of muons successfully
fitted to a common vertex. The data sample corresponds to an integrated
luminosity of 11.4~fb$^{-1}$~\cite{Aaboud:2016hhf}, collected at a proton--proton collision energy $\sqrt{s} = 8 \TeV$.
Each muon candidate reconstructed offline is required to have good spatial
matching to a trigger object, satisfying $\Delta R \equiv \sqrt{\left( \Delta\eta \right)^2+\left(\Delta\phi \right)^2} < 0.01$.
Events where two oppositely charged muon candidates are reconstructed
with pseudorapidity $|\eta^\mu|<2.3$ and transverse momenta
$\pt^\mu>4 \GeV$
are kept for further analysis only if the invariant mass of the dimuon system
falls within $\pm120\, \MeV$ of the mass of the $\jpsi$ meson, $m({J/\psi})=3096.916\pm 0.011$~MeV~\cite{PDGBcTau}.

The two muon tracks are fitted to a common vertex with a loose cut on fit quality, $\chi^2 < 200$. The dimuon invariant mass is then constrained to the $\jpsi$ mass, and the four-track vertex fit of the two muon tracks and pairs of non-muon tracks is performed to find $\jpsi\pi^+\pi^-$ candidates. The two non-muon tracks are assigned pion
masses, and are required to have opposite charges and to satisfy the conditions
$\pt^\pi > $~0.6~GeV, $|\eta^\pi| <  2.4 $.
Four-track candidates with fit $\chi^2$ probability $P(\chi^2) <  4\%$ are discarded.

Only $J/\psi \pi^{+}\pi^{-}$ combinations with rapidity $y$ within the range
$|y|<0.75$ are considered in this analysis, with most of the contributing tracks
measured within the barrel part of the detector $|\eta| \lesssim 1$ where the tracking resolution
is optimal.
Then the transverse momenta of the $J/\psi \pi^{+}\pi^{-}$ candidates are required to be within the range $10$~GeV$ < \pt < 70$~GeV.

Further selection requirements are applied to the remaining $J/\psi \pi^{+}\pi^{-}$
combinations:
\begin{equation}
\Delta R(\jpsi, \pi^{\pm}) < 0.5, \ \ \ \ \ \ \ \ \ Q < 0.3  \GeV,
\label{backgroundCuts}
\end{equation}
where $\Delta R(\jpsi, \pi^{\pm}) $ is the angular
distance between the momenta of the dimuon system and each pion candidate,
while
$Q \equiv m(\jpsi\pi^{+}\pi^{-}) - m(J/\psi) - m(\pi^{+}\pi^{-})$.
Here $m(\jpsi\pi^{+}\pi^{-})$  and
$m(\pi^{+}\pi^{-})$ are the fitted invariant masses of the ${\mu^{+}\mu^{-}\pi^{+}\pi^{-}}$
and the dipion system, respectively.
These requirements are found to be $>90 \%$ efficient for the signal
from $\psi(2S)$ and $X(3872)$ decays, while significantly suppressing the combinatorial background.

\begin{figure}[H]
\begin{center}
\subfigure[]{\includegraphics[width=0.49\textwidth]{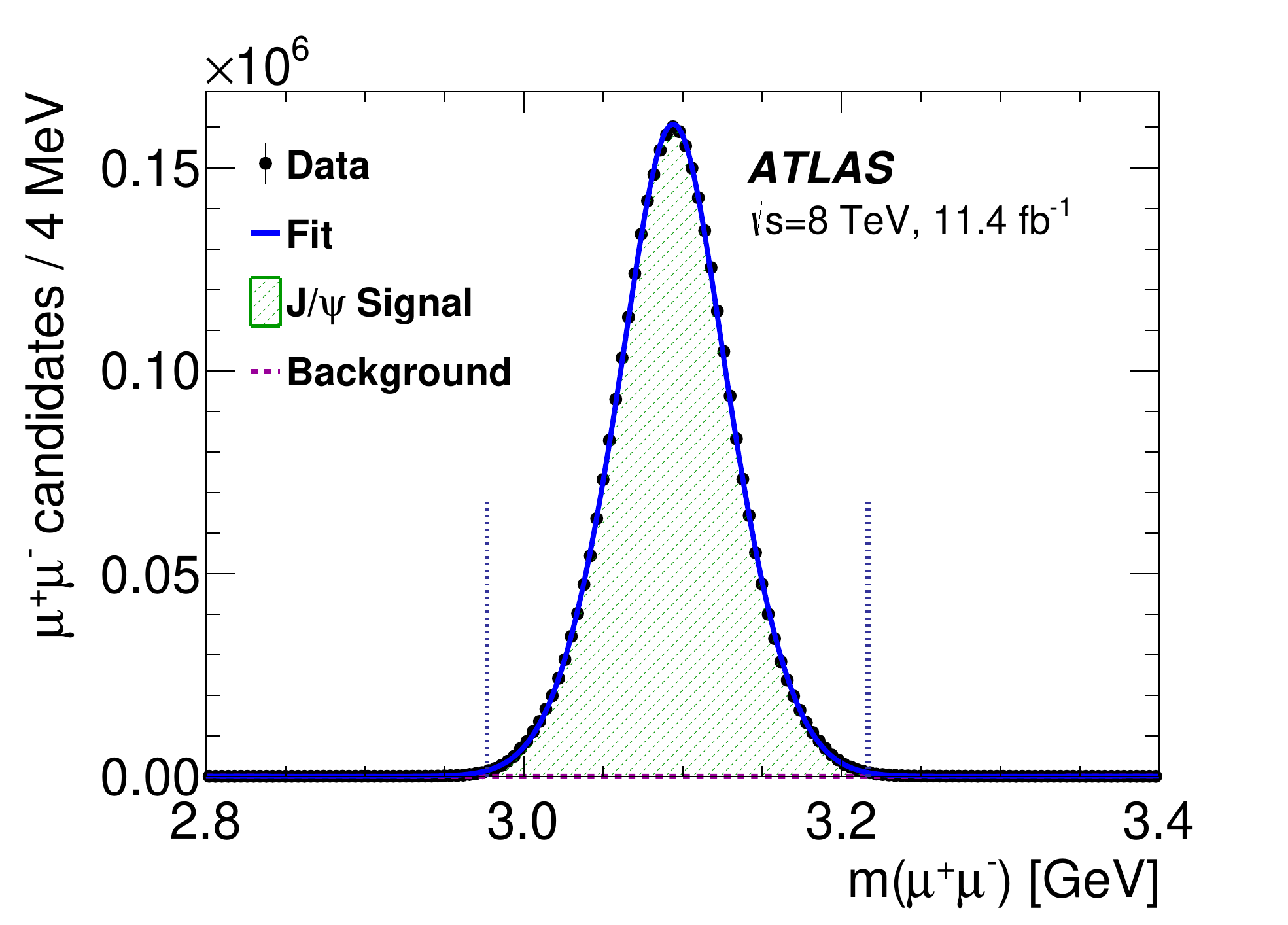}\label{jpsiYield}
}
\subfigure[]{\includegraphics[width=0.49\textwidth]{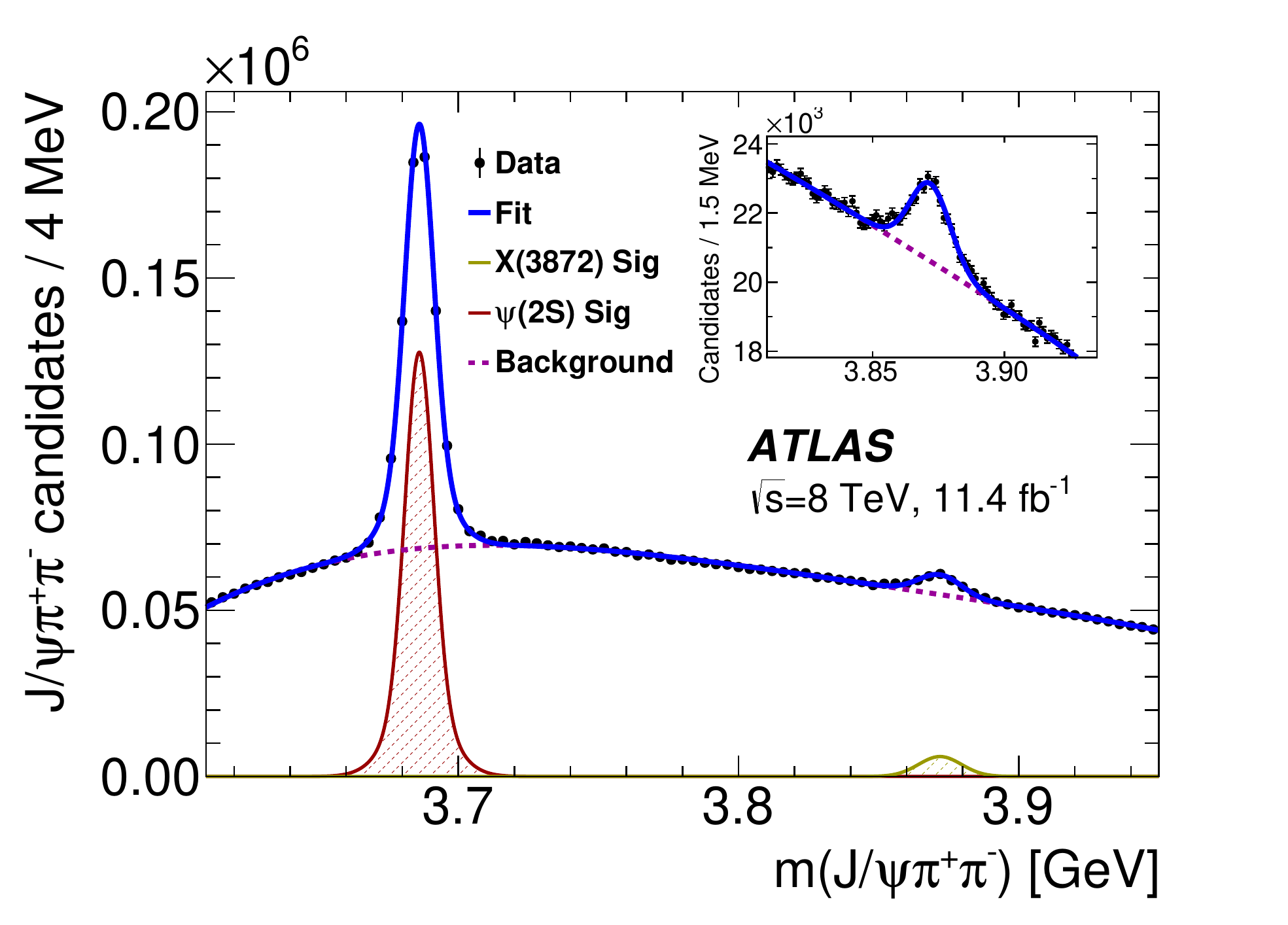}\label{totalYield}
}
\caption{\subref{jpsiYield} The invariant mass distribution of the $J/\psi$ candidates satisfying all selection criteria except the $\pm 120 \MeV$ $\jpsi$ mass window
requirement indicated here by the dotted vertical lines. The curve shows the result of a fit with a double-Gaussian function for signal and a second-order polynomial for background. \subref{totalYield} Invariant mass of the selected $J/\psi \pi^{+}\pi^{-}$ candidates collected over the full $\pt$ range 10--70~GeV and the rapidity range $|y|<0.75$ after selection requirements.
The curve shows the results of the fit using double-Gaussian functions for the
$\psi(2S)$ and $X(3872)$ peaks and a fourth-order polynomial for the background. The $X(3872)$ mass range is highlighted in the inset. }
\label{yields}
\end{center}
\end{figure}

The invariant mass distribution of the dimuons contributing to
the selected $J/\psi \pi^{+}\pi^{-}$ combinations is shown in Figure~\ref{jpsiYield}
between the dashed vertical lines. The distribution is fitted with the sum of a second-order polynomial background and a double-Gaussian function, which  contains about 3.6$\,$M $J/\psi$ candidates.
The invariant mass distribution of the $J/\psi \pi^{+}\pi^{-}$
candidates selected for further analysis is presented in Figure~\ref{totalYield}.
The fitted function is the sum of a fourth-order polynomial background and two double-Gaussian functions.
The double-Gaussian functions for $\psi(2S)$ and $X(3872)$ contain about 470$\,$k and 30$\,$k candidates, respectively.

Monte Carlo (MC) simulation is used to study the selection and reconstruction efficiencies.
The MC samples with $b$-hadron production and decays are generated with \PYTHIA$\,$6.4~\cite{Sjostrand:2006za},
complemented, where necessary, with a dedicated extension for $B_c$ production based on calculations from Refs.~\cite{Berezhnoy:1995au,Berezhnoy:1996ks,Berezhnoy:1997fp,Berezhnoy:2004gc}. The decays of $b$-hadrons are then simulated with EvtGen~\cite{Lange:2001uf}.
The generated events are passed through a full simulation of the detector using the ATLAS simulation framework~\cite{SOFT-2010-01} based on \GEANT\hspace{-3pt}$4$~\cite{Agostinelli:2002hh,Allison:2006ve} and processed with the same software as that used for the data.

\section{Analysis method}
\label{sec:crossSection}

The production cross sections of the $\psi(2S)$ and $X(3872)$ states
decaying to $J/\psi \pi^{+}\pi^{-}$ are
measured in five bins of $J/\psi \pi^{+}\pi^{-}$ transverse momentum, with bin boundaries
$(10,12,16,22,40,70) \, \GeV$.

The selected $J/\psi \pi^{+}\pi^{-}$ candidates are weighted
in order to correct for signal loss
at various stages of the selection process.
Following previous similar analyses
\cite{BPHY-2013-06,BPHY-2012-02b}
a per-candidate weight $\omega$ was calculated as
\begin{equation}
\small
\omega = \left[
{\cal A}(\pt,y)
\cdot \epsilon_{\mathrm{trig}}(\pt^{\mu^{\pm}}, \eta^{\mu^{\pm}},y^{\jpsi})
\cdot\epsilon^{\mu}(\pt^{\mu+}, \eta^{\mu+})
\cdot\epsilon^{\mu}(\pt^{\mu-}, \eta^{\mu-})
\cdot\epsilon^{\pi}(\pt^{\pi+}, \eta^{\pi+})
\cdot\epsilon^{\pi}(\pt^{\pi-}, \eta^{\pi-})
\right]^{-1}.
\end{equation}
Here, $\pt$ and $y$ stand for the transverse momentum and rapidity of the $\jpsi\pi^+\pi^-$ candidate, $y^{\jpsi}$ is the rapidity of the $\jpsi$ candidate, while $\pt^{\pi^{\pm}}, \pt^{\mu^\pm}, \eta^{\pi^\pm}$ and $\eta^{\mu^\pm}$ are transverse momenta and pseudorapidities of the respective pions and muons. The trigger efficiency $\epsilon_{\mathrm{trig}}$ and the muon reconstruction
efficiency $\epsilon^{\mu}$ were obtained using data-driven tag-and-probe methods described in~Refs.~\cite{BPHY-2012-02b,muonID}.
The pion reconstruction efficiency $\epsilon^{\pi}$ is obtained through MC
simulations using the method described in~Ref.~\cite{BPHY-2013-06}.

The acceptance ${\cal A}(\pt,y)$ is defined as the probability that
the muons and pions comprising a $J/\psi \pi^{+}\pi^{-}$ candidate with
transverse momentum $\pt$ and rapidity $y$ fall within the fiducial limits
described in Section~\ref{sec:eventsel}. The acceptance map is created using
generator-level simulation, with small reconstruction-level corrections
applied at a later stage
(see~Ref.~\cite{BPHY-2012-02b} for more details).
The different quantum numbers of the $\psi(2S)$ and $X(3872)$ ($J^{PC} =
1^{--}$ and $1^{++}$, respectively) cause a difference in the expected
dependence of the acceptance on the spin-alignments of the two states. The cross
sections measured in this paper are obtained assuming no spin-alignment,
but appropriate sets of correction factors for a number of extreme spin-alignment scenarios are calculated and presented in Appendix~\ref{app:spinAllignment} for each $\pt$ bin, separately for $\psi(2S)$ and $X(3872)$.

The efficiencies of the reconstruction-quality requirements and the background-suppression requirements described in Section~\ref{sec:eventsel} are determined using MC simulations, and the corrections are applied in each of the $\pt$ bins, separately for $\psi(2S)$ and $X(3872)$. These efficiencies are found to vary between $84\%$ and $95\%$. The simulated distributions are reweighted to match the data,
and values with and without reweighting are used to estimate systematic uncertainties (see Section~\ref{sec:systematics}).

In order to separate prompt production of the
$\psi(2S)$ and $X(3872)$
states from
the non-prompt production occurring via the decays of long-lived particles such as $b$-hadrons, the data sample in each
$\pt$ bin is
further divided into intervals of pseudo-proper lifetime $\tau$, defined as
\begin{equation}
\tau = \frac{L_{xy}\hspace{1pt} m}{c \pt},
\end{equation} where $m$ is the invariant mass,
$\pt$ is the transverse momentum and $L_{xy}$ is the transverse decay length
of the $J/\psi \pi^{+}\pi^{-}$ candidate. $L_{xy}$ is defined as
\begin{equation}
L_{xy}= \frac{\vec{L} \cdot \vec{p}_{\mathrm{T}}}{\pt}, 
\end{equation}
where $\vec{L}$ is the vector pointing from the primary $pp$ collision vertex
to the $J/\psi \pi^{+}\pi^{-}$ vertex, while $\vec{p}_{T}$ is the transverse
momentum vector of the $J/\psi \pi^{+}\pi^{-}$ system. The coordinates of the primary vertices (PV) are obtained from charged-particle tracks with $\pt > 0.4$~GeV not used in the decay vertices, and are transversely constrained to the luminous region of the colliding beams.
The matching of a $J/\psi \pi^{+}\pi^{-}$ candidate to a PV is made by finding the one with the smallest three-dimensional impact parameter, calculated between the $J/\psi \pi^{+}\pi^{-}$ momentum and each PV.

Based on an analysis of the lifetime
resolution and lifetime dependence of the signal, four lifetime intervals were
defined:
\begin{itemize}
\item[\\\\ $w_0: $] \hspace{15pt}$-0.3 \hspace{2pt} \mathrm{ps} < \tau(J/\psi\pi\pi) < 0.025
\hspace{2pt} \mathrm{ps}$,
\item[\\\\ $w_1: $] \hspace{15pt}$0.025 \hspace{2pt} \mathrm{ps} < \tau(J/\psi\pi\pi) < 0.3
\hspace{2pt} \mathrm{ps}$,
\item[\\\\ $w_2 :$] \hspace{15pt}$0.3 \hspace{2pt} \mathrm{ps} < \tau(J/\psi\pi\pi) < 1.5
\hspace{2pt} \mathrm{ps}$,
\item[\\\\ $w_3: $] \hspace{15pt}$1.5 \hspace{2pt} \mathrm{ps} < \tau(J/\psi\pi\pi) < 15.0
\hspace{2pt} \mathrm{ps}$.
\end{itemize}

\begin{figure}
\begin{center}
\subfigure[]{\includegraphics[width=0.49\textwidth]{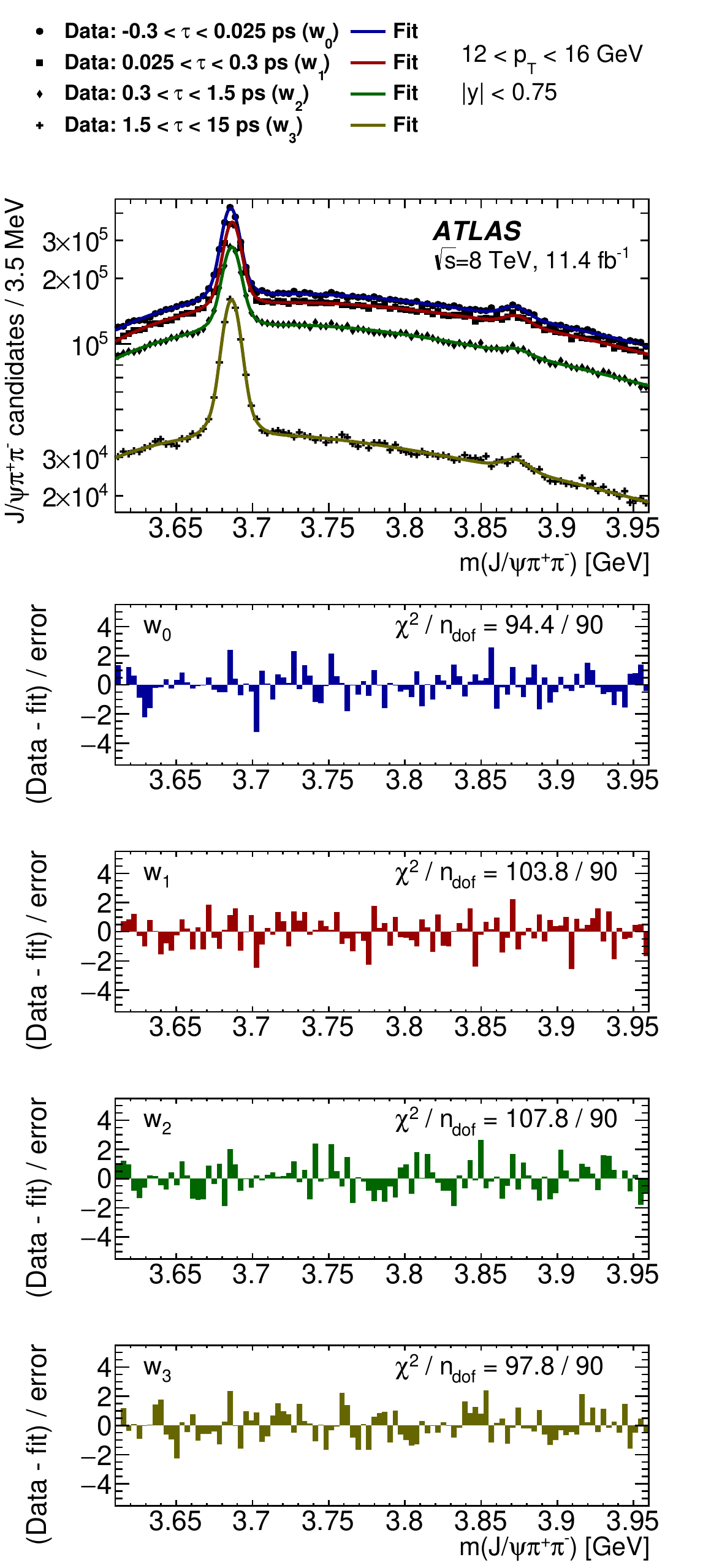}\label{mFit1} }
\subfigure[]{\includegraphics[width=0.49\textwidth]{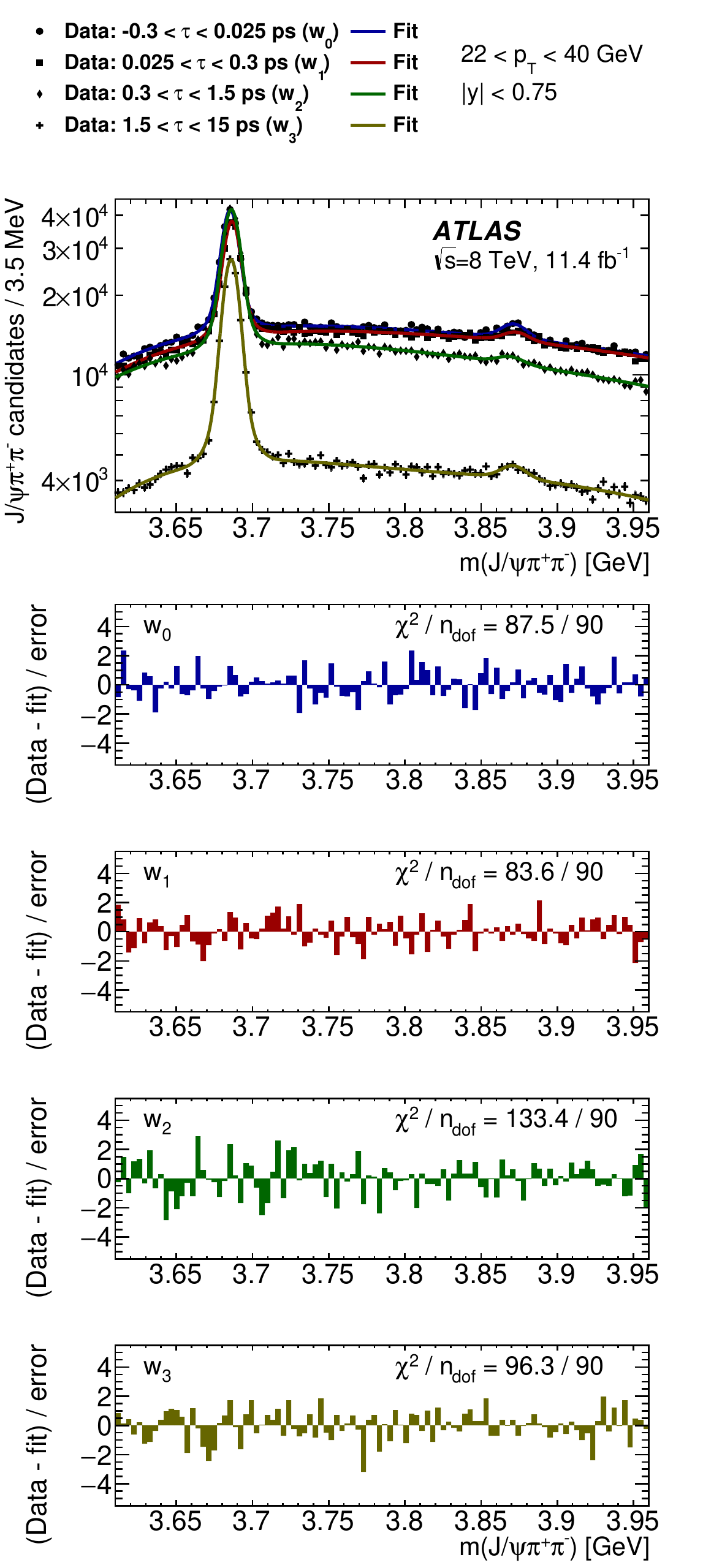}\label{mFit2} }
\caption{ The invariant mass spectra of the $\jpsi\pi^+\pi^-$ candidates to extract $\psi(2S)$ and $X(3872)$ signal for each pseudo-proper lifetime window in the $\pt$ bin \subref{mFit1} $[12,16] \GeV$ and \subref{mFit2} $[22,40] \GeV$. Shown underneath the fits are the corresponding pull distributions,
with respective values of $\chi^2$ per degree of freedom for each fit. }
\label{mFit12}
\end{center}
\end{figure}

In each of these intervals, and for each $\pt$ bin, the invariant mass
distribution
of the $J/\psi \pi^{+}\pi^{-}$ system is built using fully corrected weighted
events.
These distributions are shown in Figure \ref{mFit12} for representative $\pt$ bins.

In order to determine the yields of the $\psi(2S)$ and $X(3872)$ signals, the
distributions are fitted in each lifetime interval to the function:
\begin{equation}
\begin{split}
f(m) =
Y^{\psi}
\left(f_{1} G^{\psi}_1(m) + \left( 1-f_{1}\right) G^{\psi}_2(m)\right)
+ Y^{X}
\left(f_{1} G^{X}_1(m) + \left( 1-f_{1}\right) G^{X}_2(m)\right)
\\
+ N (m-m_{\mathrm{th}})^{p_1}
\mathrm{e}^{p_{2}(m-m_{\mathrm{th}})} P(m-m_{\mathrm{th}}),
\end{split}
\label{mainFitEqn}
\end{equation}
where the threshold mass $m_{\mathrm{th}} = m_{J/\psi} + 2 m_\pi = 3376.06$~MeV.
The $\psi(2S)$ and $X(3872)$ signal yields $Y^{\psi}$ and  $Y^{X}$,
coefficients of the second-order polynomial $P$, parameters $p_1$ and $p_2$, and the normalisation of the background term $N$, are determined from the fits.
Signal peaks for $\psi(2S)$ and $X(3872)$ are described by normalised double-Gaussian functions with common means: $G^{\psi}_{1}(m)$ and $G^{X}_{1}(m)$ are the
narrower Gaussian functions with respective widths $\sigma_\psi$ and $\sigma_X$,
while $G^{\psi}_{2}(m)$ and $G^{X}_{2}(m)$ are wider Gaussian functions with
widths $2\sigma_\psi$ and $2\sigma_X$.
The fraction of the narrower Gaussian function $f_{1}$ is assumed to be the same for
$\psi(2S)$ and $X(3872)$, while the widths $\sigma_\psi$ and $\sigma_X$
are related by $\sigma_X = \kappa \sigma_\psi$. The parameters
$f_{1}$ and $\kappa$ are fixed for the main fits to the values $f_{1} = 0.76
\pm 0.04$, $\kappa = 1.52 \pm 0.05$ as determined from a fit applied in the range 16~GeV$< \pt < 70$~GeV, which offers a better signal-to-background ratio than the full range, and is varied within these errors in the systematic uncertainty studies.
The fit quality is found to be good throughout the range of
transverse momenta and lifetimes.
The yields extracted from the fits are shown in Table~\ref{tab:massfitValues_psi2s_rap} for the $\psi(2S)$ and Table~\ref{tab:massfitValues_X3872_rap} for the $X(3872)$.

\begin{table}[htbp]
\renewcommand{\arraystretch}{1.15}
\begin{center}
\begin{tabular}{c r@{\hspace{1mm}$\pm$\hspace{1mm}}l r@{\hspace{1mm}$\pm$\hspace{1mm}}l r@{\hspace{1mm}$\pm$\hspace{1mm}}l r@{\hspace{1mm}$\pm$\hspace{1mm}}l r@{\hspace{1mm}$\pm$\hspace{1mm}}l}
\hline\hline
& \multicolumn{10}{c}{Corrected yields of $\psi(2\mathrm{S})$ $[\times 10^{5}]$  vs. $p_{\rm T}$ [GeV]} \\
$\tau$ window  &\multicolumn{2}{c}{10--12} &  \multicolumn{2}{c}{12--16}  &  \multicolumn{2}{c}{16--22}  &  \multicolumn{2}{c}{22--40} &  \multicolumn{2}{c}{40--70}  \\ \hline
$w_{0}$ & $ 17.48$ &  $0.36$ & $ 11.03$ &  $0.11$ & $3.53$ &  $0.03$ & $1.14$ &  $0.01$ & $0.078$ &  $0.004$ \\
$w_{1}$ & $ 14.07$ &  $0.37$ & $ 9.04$ &  $0.10$ & $2.94$ &  $0.03$ & $1.01$ &  $0.01$ & $0.071$ &  $0.003$ \\
$w_{2}$ & $ 9.13$ &  $0.29$ & $ 7.04$ &  $0.09$ & $2.97$ &  $0.03$ & $1.27$ &  $0.01$ & $0.104$ &  $0.004$ \\
$w_{3}$ & $ 6.74$ &  $0.16$ & $ 5.21$ &  $0.06$ & $2.22$ &  $0.02$ & $0.94$ &  $0.01$ & $0.081$ &  $0.003$ \\
\hline\hline
\end{tabular}
\end{center}
\caption{\label{tab:massfitValues_psi2s_rap}Fitted yields of $\psi(2\mathrm{S})$ in bins of pseudo-proper lifetime and $p_{\rm T}$. Uncertainties are statistical only.}
\end{table}

\begin{table}[htbp]
\renewcommand{\arraystretch}{1.15}
\begin{center}
\begin{tabular}{c r@{\hspace{1mm}$\pm$\hspace{1mm}}l r@{\hspace{1mm}$\pm$\hspace{1mm}}l r@{\hspace{1mm}$\pm$\hspace{1mm}}l r@{\hspace{1mm}$\pm$\hspace{1mm}}l r@{\hspace{1mm}$\pm$\hspace{1mm}}l}
\hline\hline
& \multicolumn{10}{c}{Corrected yields of $X(3872)$ $[\times 10^{4}]$  vs. $p_{\rm T}$ [GeV]} \\
$\tau$ window  &\multicolumn{2}{c}{10--12} &  \multicolumn{2}{c}{12--16}  &  \multicolumn{2}{c}{16--22}  &  \multicolumn{2}{c}{22--40} &  \multicolumn{2}{c}{40--70}  \\ \hline
$w_{0}$ & $ 10.8$ &  $2.3$ & $ 10.55$ &  $0.76$ & $3.53$ &  $0.26$ & $1.19$ &  $0.11$ & $0.093$ &  $0.030$ \\
$w_{1}$ & $ 9.3$ &  $2.7$ & $ 8.21$ &  $0.71$ & $2.60$ &  $0.24$ & $0.72$ &  $0.11$ & $0.039$ &  $0.023$ \\
$w_{2}$ & $ 4.1$ &  $1.7$ & $ 3.83$ &  $0.63$ & $1.29$ &  $0.21$ & $0.45$ &  $0.10$ & $0.036$ &  $0.023$ \\
$w_{3}$ & $ 2.06$ &  $0.81$ & $ 2.09$ &  $0.34$ & $0.98$ &  $0.13$ & $0.30$ &  $0.06$ & $0.020$ &  $0.014$ \\
\hline\hline
\end{tabular}
\end{center}
\caption{\label{tab:massfitValues_X3872_rap}Fitted yields of $X(3872)$ in bins of pseudo-proper lifetime and $p_{\rm T}$. Uncertainties are statistical only.}
\end{table}

Once the corrected yields $Y^{\psi}$ and $Y^{X}$ are determined in each
$\pt$ bin, the double differential cross sections
(times the product of the relevant branching fractions)
can be calculated:
\begin{equation}
\mathcal{B}(i \rightarrow J/\psi \pi^{+}\pi^{-} )
\mathcal{B}( J/\psi \rightarrow \mu^{+}\mu^{-} )
\frac{\mathrm{d}^2\sigma(i)}{\mathrm{d}\pt dy}  = \frac{Y^{i}}{\Delta \pt\Delta y
\int{\hspace{-3pt}\cal{L}}dt },
\end{equation}
where $i$ stands for $\psi(2S)$ or $X(3872)$, $\int{\hspace{-3pt}\cal{L}}dt $ is the integrated luminosity, while $\Delta \pt$ and  $\Delta y$ are widths of the
relevant transverse momentum and rapidity bins, with $\Delta y=1.5$.
$\mathcal{B}(i\rightarrow J/\psi \pi^{+}\pi^{-} )$ and $\mathcal{B}( J/\psi
\rightarrow \mu^{+}\mu^{-})$ are the branching fractions of these respective
decays.

\section{Lifetime fits}
\label{sec:LifetimeFits}
The probability density function (PDF) describing the dependence of
$\psi(2S)$ and $X(3872)$ signal yields on the pseudo-proper lifetime $\tau$ is
a superposition of prompt (P) and non-prompt
(NP) components:
\begin{equation}
F^i(\tau) = (1-f^i_{\mathrm{NP}}) F^i_{\mathrm{P}}(\tau) + f^i_{\mathrm{NP}} F^i_{\mathrm{NP}}(\tau),
\label{mainLifetimeEq}
\end{equation}
where $f_{\mathrm{NP}}$ is the non-prompt fraction, while $i$ stands for either $\psi(2S)$ or $X(3872)$.
The prompt components of $\psi(2S)$ and $X(3872)$ production should not have
any observable decay length, and hence $F_{\mathrm{P}}(\tau)$ is effectively described by
the lifetime resolution function $F_{\mathrm{res}}(\tau)$, assumed to be the
same for
$\psi(2S)$ and $X(3872)$ signals. This was verified with simulated data samples.
The resolution function $F_{\mathrm{res}}(\tau)$ is parameterised as a
weighted sum of three normalised Gaussian functions with a common mean, with respective
width parameters $\sigma_1=\sigma_\tau$, $\sigma_2=2\sigma_\tau $ and
$\sigma_3=4\sigma_\tau $.
The resolution parameter $\sigma_{\tau}$ and the relative weights of the three Gaussian functions are determined separately for each analysis
$\pt$ bin, using two-dimensional mass--lifetime unbinned maximum-likelihood fits
on the subset of data which contains a narrow range of masses around the
$\psi(2S)$ peak.
The fitted values for $\sigma_{\tau}$ are within the range of 32--52~fs, with
the weight of the narrowest Gaussian function steadily increasing with $\pt$ from 6$\%$
to about 50$\%$.

\begin{figure}[h]
\begin{center}
\subfigure[]{\includegraphics[width=0.49\textwidth]{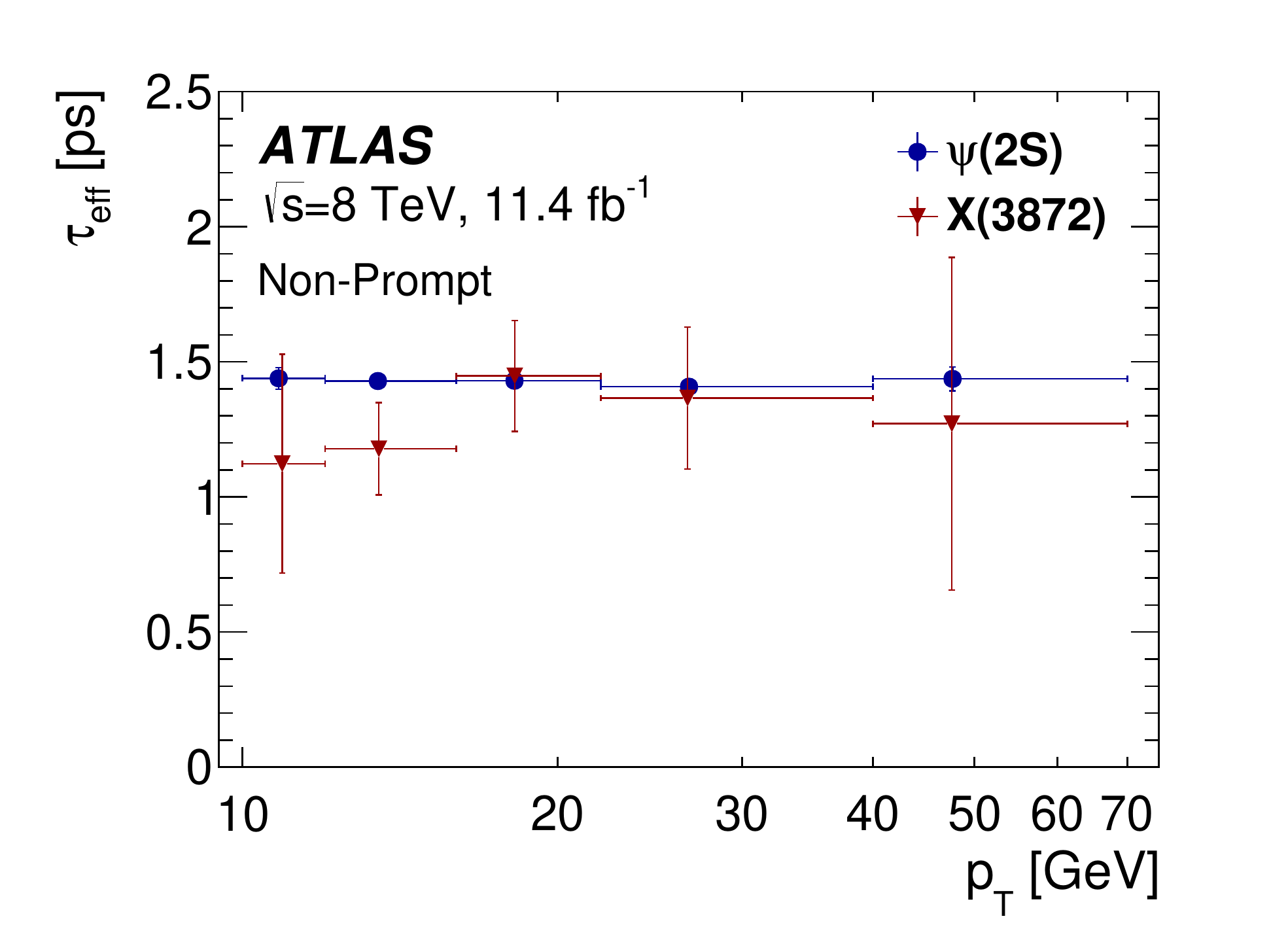}\label{fig:lifetimes} }
\subfigure[]{\includegraphics[width=0.49\textwidth]{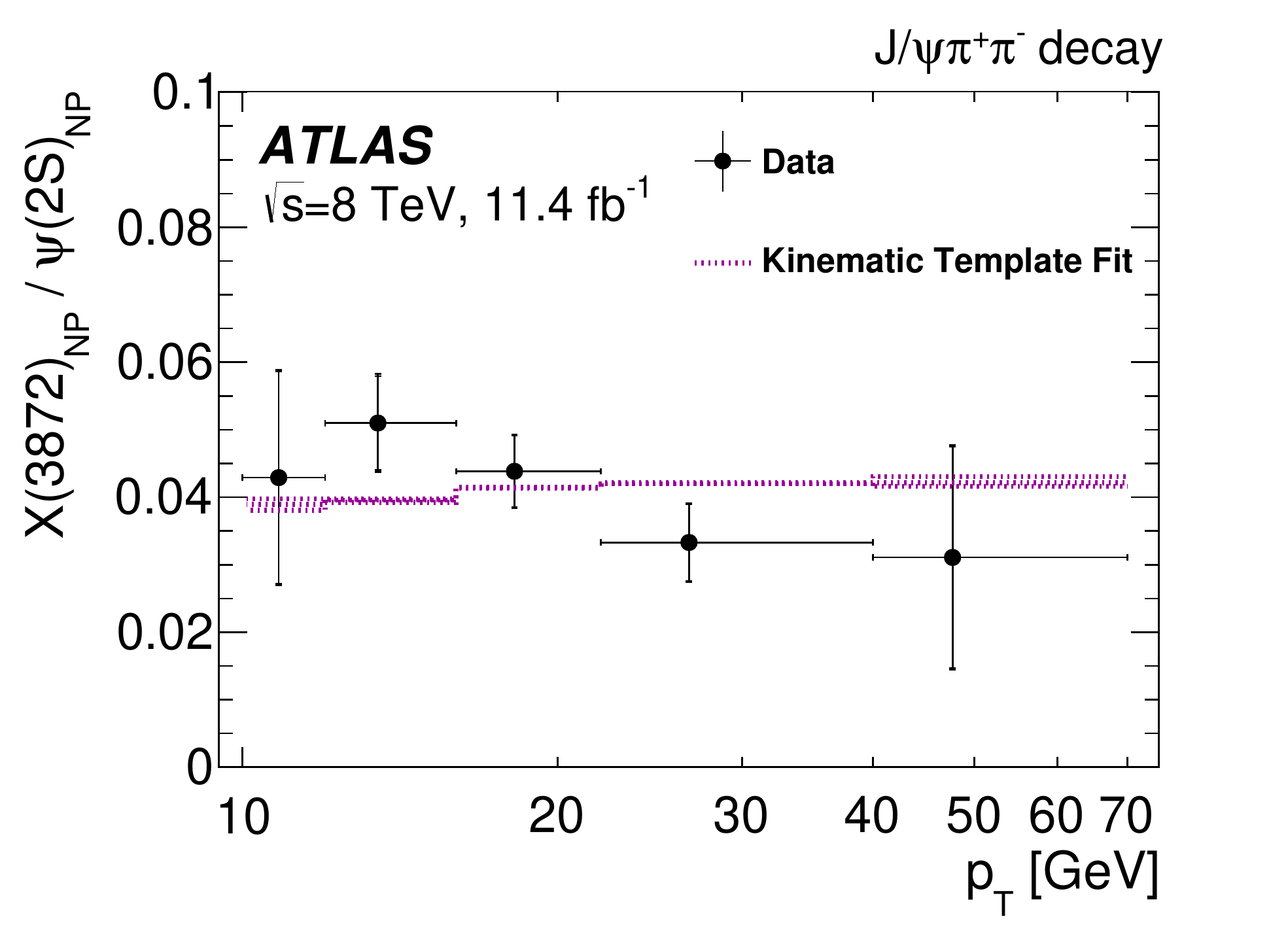}\label{fig:npr} }
\caption{\subref{fig:lifetimes} Measured effective pseudo-proper lifetimes for non-prompt $X(3872)$ and $\psi(2S)$. \subref{fig:npr} Ratio of non-prompt production cross sections times branching fractions, $X(3872) / \psi(2S)$, in the single-lifetime fit model.
The measured distribution is fitted to the kinematic template described in the text.}
\label{fig:lifetimesNpr}
\end{center}
\end{figure}

The simplest description of the non-prompt components of the signal PDF is
given by a single one-sided exponential smeared with the resolution function,
with the effective lifetime $\tau_{\mathrm{eff}}$
determined from the fit. This model, referred to as a `single-lifetime fit',
is applied to the
$\psi(2S)$ and $X(3872)$ yields from Tables~\ref{tab:massfitValues_psi2s_rap} and \ref{tab:massfitValues_X3872_rap}, and the results of the
corresponding binned minimum-$\chi^2$ fits are shown in Figure \ref{fig:lifetimesNpr}.

\begin{table}[h]
\renewcommand{\arraystretch}{1.1}
\begin{center}
\begin{tabular}{c|c|c} \hline \hline
$\pt$ bin [GeV]	& $\tau_{\mathrm{eff}}(\psi(2S))$ [ps] 				& $\tau_{\mathrm{eff}}(X(3872))$ [ps] 			 \\ \hline
10--12 				& $1.44\pm0.04$ 		& $1.12\pm0.40$  \\
12--16 				& $1.43\pm0.02$ 		& $1.18\pm0.17$  \\
16--22 				& $1.43\pm0.01$ 		& $1.45\pm0.21$  \\
22--40 				& $1.41\pm0.01$ 		& $1.37\pm0.26$  \\
40--70				& $1.44\pm0.04$ 		& $1.27\pm0.62$  \\ \hline \hline
\end{tabular}
\caption{Effective pseudo-proper lifetimes for non-prompt $\psi(2S)$ and $X(3872)$ obtained with the single-lifetime fit model.}
\label{tauStudyTableAltSingle}
\end{center}
\end{table}

Figure~\ref{fig:lifetimes} shows the effective pseudo-proper lifetimes $\tau_{\mathrm{eff}}$
for non-prompt $\psi(2S)$ and $X(3872)$ signals in bins of $\pt$ (see also Table~\ref{tauStudyTableAltSingle}).
While for $\psi(2S)$ the fitted values of  $\tau_{\mathrm{eff}}$ are measured to be around $1.45$~ps in all $\pt$ bins,
the signal from $X(3872)$ at low $\pt$ tends to have shorter
lifetimes, possibly hinting at a different production mechanism at low \pt.

In Figure~\ref{fig:npr} the ratio of non-prompt production cross sections of $X(3872)$ and $\psi(2S)$, times respective branching fractions, for the single-lifetime fit is plotted as a
function of transverse momentum. The measured distribution is compared to the
kinematic template,  which is calculated as a  ratio of the simulated $\pt$ distributions of non-prompt $X(3872)$ and non-prompt $\psi(2S)$, assuming that the same mix of the parent $b$-hadrons contributes to both signals.
The shape of the template reflects the kinematics of the decay of a $b$-hadron
into $\psi(2S)$ or $X(3872)$, with the width of the band showing the range of variation for extreme values of the invariant mass of the recoiling hadronic system.
A fit of the measured ratio to this template allows determination of the ratio of the average branching fractions:
\begin{equation}
R^{\mathrm{1L}}_{B} =  \frac{\mathcal{B}(B \rightarrow X(3872) \textrm{ + any})  \mathcal{B}(X(3872) \rightarrow \jpsi\pi^+\pi^-)}{\mathcal{B}(B \rightarrow \psi(2S) \textrm{ + any})  \mathcal{B}(\psi(2S) \rightarrow \jpsi\pi^+\pi^-)} = (3.95 \pm 0.32 \mathrm{(stat)} \pm 0.08\mathrm{(sys)}) \times 10^{-2},
\label{rb.1l}
\end{equation}
where the systematic uncertainty reflects the variation of the kinematic template. The $\chi^2$ of the fit is 5.4 for the four degrees of freedom (dof).

An alternative lifetime model, also implemented in this analysis, allows for two non-prompt contributions with distinctly
different effective lifetimes (the `two-lifetime fit').
The statistical power of the
data sample is insufficient for determining two free lifetimes, especially in
the case of  $X(3872)$ production, so in this fit model the non-prompt PDFs are
represented in each $\pt$ bin by a sum of two contributions with different fixed lifetimes, and
a relative weight determined by the fit:
\begin{equation}
F^i_{\mathrm{NP}}(\tau) = (1-f^i_{\mathrm{SL}}) F_{\mathrm{LL}}(\tau) + f^i_{\mathrm{SL}} F_{\mathrm{SL}}(\tau).
\label{mainLifetimeEq2}
\end{equation}
Here, the labels SL and LL refer to short-lived and long-lived non-prompt
components, respectively,
and $f^i_{\mathrm{SL}}$ are the short-lived non-prompt fractions for $i=\psi(2S), X(3872)$.
The PDFs $F_{\mathrm{SL}}(\tau)$ and $F_{\mathrm{LL}}(\tau)$ are parameterised as single one-sided exponential functions with fixed lifetimes, smeared with the lifetime resolution function
$F_{\mathrm{res}}(\tau)$ described above.
Any long-lived part of the non-prompt contribution is assumed to originate from the usual mix of $B^\pm, B^0, B_s$ mesons and $b$-baryons, while any short-lived part would be due to the contribution of $B_c^{\pm}$ mesons.

Simulations show that the observed effective pseudo-proper lifetime of $\psi(2S)$ or $X(3872)$ from $B_c$ decays depends on the invariant mass
of the hadronic system recoiling from the hidden-charm state.
Within the kinematic range of this measurement, it varies from about $0.3$~ps for
small masses of the recoiling system to about $0.5$~ps for the largest ones. The majority of the decays are expected to
have masses of the recoiling system between these values, therefore $\tau_{\mathrm{SL}}$ is taken as the mean of the two extremes, $0.40\pm 0.05$~ps.

The effective pseudo-proper lifetime of the long-lived component, $\tau_{\mathrm{LL}}$,
is determined from the two-lifetime test fits to the $\psi(2S)$ mass range, with $\tau_{\mathrm{LL}}$ free and allowing for an
unknown contribution of a short-lived component with lifetime $\tau_{\mathrm{SL}}$.
Across the \pt bins, $\tau_{\mathrm{LL}}$ is found to be within the range $1.45\pm 0.05$~ps.
The effective pseudo-proper lifetimes $\tau_{\mathrm{LL}}$ and $\tau_{\mathrm{SL}}$ are fixed to the above values
for the main fits, and are varied within the quoted errors during systematic uncertainty studies.

\begin{figure}[H]
\begin{center}
\subfigure[]{\includegraphics[width=0.47\textwidth]{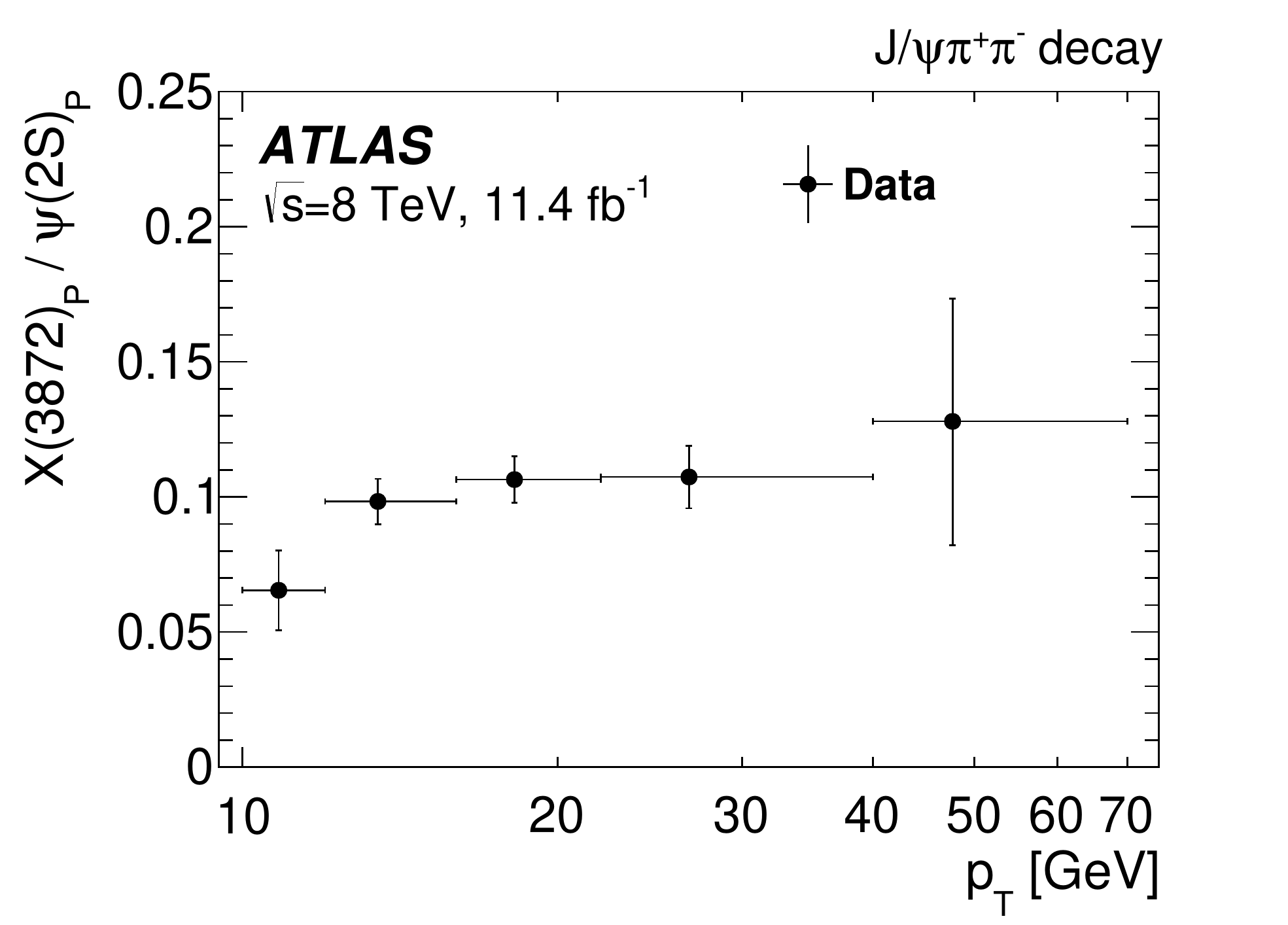}\label{fig:prompt_ratio}}
\subfigure[]{\includegraphics[width=0.47\textwidth]{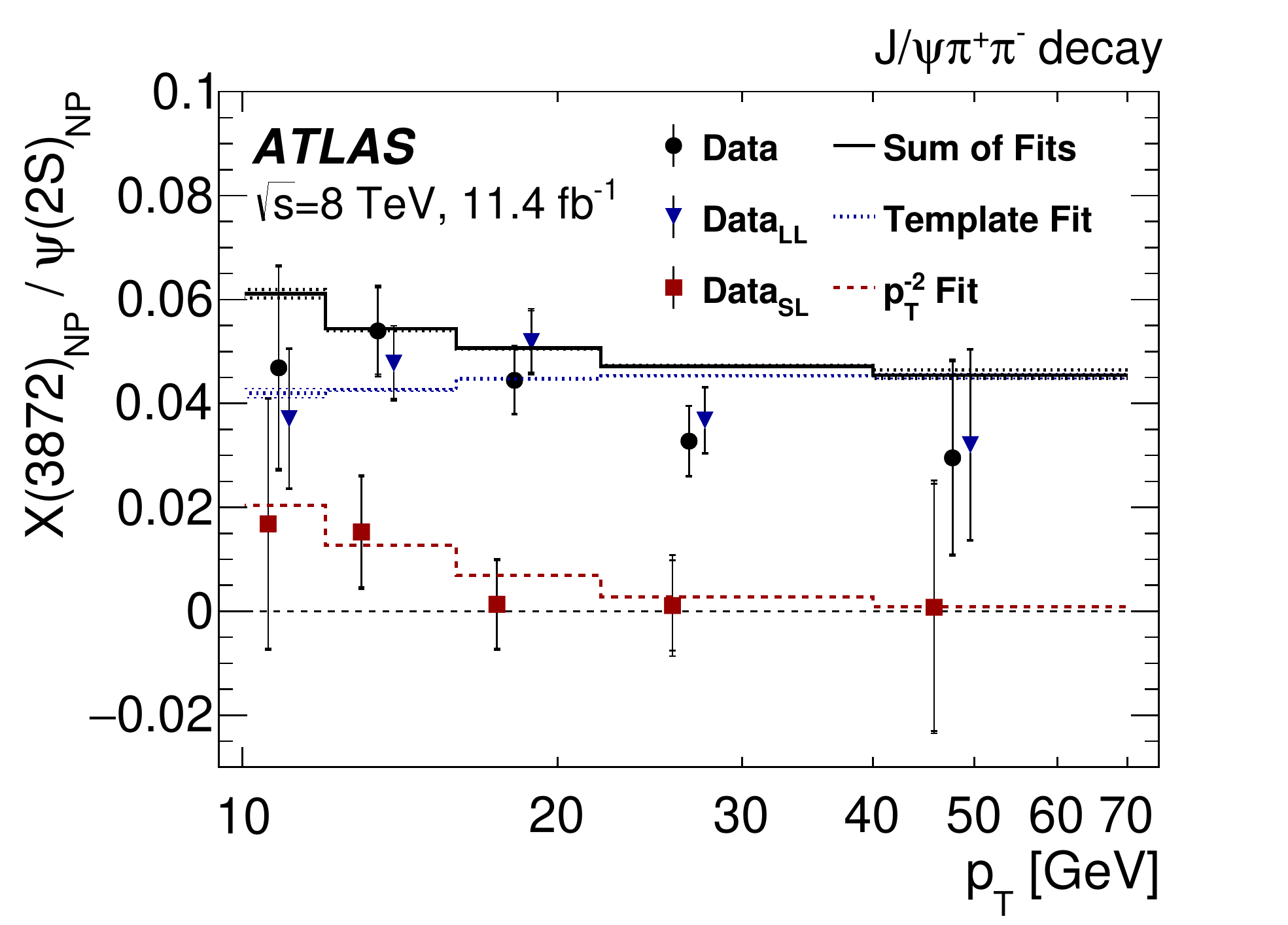}\label{fig:nonprompt_ratio}}
\caption{Ratio of cross sections times branching fractions, $X(3872) / \psi(2S)$, for \subref{fig:prompt_ratio} prompt and \subref{fig:nonprompt_ratio} non-prompt production, in the two-lifetime fit model. In \subref{fig:nonprompt_ratio}, the total non-prompt ratio (black circles) is separated into short-lived (red squares) and long-lived (blue triangles) components for the $X(3872)$, shown with respective fits described in the text.
The data points are slightly shifted horizontally for visibility.}
\label{fig:ratios}
\end{center}
\end{figure}

Figure~\ref{fig:ratios} shows the $\pt$ dependence of the ratio of $X(3872)$ to $\psi(2S)$
cross sections (times respective branching fractions), separately for prompt and non-prompt production contributions.
The non-prompt production cross section of $X(3872)$ is further split into short-lived and long-lived components. The short-lived contribution to non-prompt $\psi(2S)$ production is found to be not significant (see Table~\ref{tab:XS_ratios_TL} below). The measured ratio of long-lived $X(3872)$ to long-lived $\psi(2S)$, shown in Figure~\ref{fig:nonprompt_ratio} with blue triangles, is fitted with the MC kinematic template described before to obtain
\begin{equation}
R^{\mathrm{2L}}_{B} =  \frac{\mathcal{B}(B \rightarrow X(3872) \textrm{ + any})  \mathcal{B}(X(3872) \rightarrow \jpsi\pi^+\pi^-)}{\mathcal{B}(B \rightarrow \psi(2S) \textrm{ + any}) \mathcal{B}(\psi(2S) \rightarrow \jpsi\pi^+\pi^-)} = (3.57 \pm 0.33 \mathrm{(stat)} \pm 0.11\mathrm{(sys)}) \times 10^{-2},
\label{arbe}
\end{equation}
with $\chi^2/{\mathrm{dof}} = 2.3 / 4 $. This value of $R_{B}$ is somewhat lower than the corresponding result in
Equation~(\ref{rb.1l})
obtained from the same data with the single-lifetime fit model. Either is significantly smaller
than the value $0.18\,\pm \,0.08$ obtained by using
the estimate for the numerator, $(1.9\pm0.8) \times 10^{-4}$~\cite{Artoisenet:2009wk},
obtained from the Tevatron data,
and the world average values for the branching fractions in the denominator: $\mathcal{B}(B \rightarrow \psi(2S))=(3.07\pm0.21) \times 10^{-3}$, $ \mathcal{B}(\psi(2S) \rightarrow \jpsi\pi^+\pi^-)=(34.46\pm0.30)\%$.

Production of $B_c$ mesons in high-energy hadronic collisions at low transverse momentum is expected to be dominated by non-fragmentation processes~\cite{Berezhnoy:2013cda}.
These processes are expected to have \pt dependence $\propto \pt^{-2}$ relative to the fragmentation contribution, while it is the fragmentation contribution which dominates the production of long-lived $b$-hadrons~\cite{Cacciari:2012ny}.

So the ratio of short-lived non-prompt $X(3872)$ to non-prompt $\psi(2S)$, shown in Figure \ref{fig:nonprompt_ratio} with red squares, is fitted with a function $a/\pt^2$ to find $a = 2.04 \pm 1.43{\mathrm{(stat)}} \pm 0.34{\mathrm{(sys)}}\GeV^2$, with $\chi^2/{\mathrm{dof}} = 0.43 / 4 $. This value of $a$, and the measured non-prompt yields of $X(3872)$ and $\psi(2S)$ states, are used to determine the fraction of non-prompt $X(3872)$ from short-lived sources, integrated over the $\pt$ range ($\pt > 10 \GeV$) covered in this measurement, giving:
\begin{equation}
\frac{\sigma(pp\to B_c)\mathcal{B}(B_c\to X(3872))}{\sigma(pp\to \textrm{non-prompt\ }X(3872))} = (25 \pm 13\mathrm{(stat)} \pm 2\mathrm{(sys)}\pm 5\mathrm{(spin)})\%,
\end{equation}
where the last uncertainty comes from varying the spin-alignment of $X(3872)$ over the extreme scenarios discussed in Appendix \ref{app:spinAllignment}.
Since $B_c$ production is only small fraction of the inclusive beauty production, this value of the ratio would mean that the production of
$X(3872)$ in $B_c$ decays is strongly enhanced compared to its production in the decays of other $b$-hadrons.

The two-lifetime fits are used for $\psi(2S)$ and $X(3872)$
to obtain all subsequent results in this paper, unless specified otherwise,
with the relatively small differences between the results of the single-lifetime and two-lifetime fits
being highlighted alongside all other sources of systematic uncertainty.

\section{Systematic uncertainties}
\label{sec:systematics}

The sources of various uncertainties and their smallest (Min), median (Med) and largest (Max)
values across the $\pt$ bins are summarised in Table~\ref{tab:sysErrXS} for
the differential cross sections of $X(3872)$ and $\psi(2S)$ states,
and in Table~\ref{tab:sysErrFrac} for the measured fractions.
\begin{table}[h]
\renewcommand{\arraystretch}{1.1}
\begin{center}
\begin{tabular}{l||rrr|rrr}
\hline
& \multicolumn{3}{c}{$\psi(2S) [\%]$ } &  \multicolumn{3}{c}{$X(3872) [\%]$ } \\
Source of uncertainty 						& Min & Med & Max & Min & Med & Max \\
\hline \hline
Statistical									    	& $0.9$ & $1.4$ & $5.4$ & $7.3$ & $9.9$ & $63$\hspace{1.5mm} \\
\hline
Trigger eff. 					& $1.0$ & $1.3$ & $2.5$ & $1.1$ & $1.3$ & $2.6$ \\
Muon tracking 	& $2.0$ & $2.0$ & $2.0$ & $2.0$ & $2.0$ & $2.0$ \\
Muon reconstruction eff. 		& $0.2$ & $0.2$ & $0.3$ & $0.2$ & $0.2$ & $0.4$ \\
Pion reconstruction eff.	    	& $2.5$ & $2.5$ & $2.5$ & $2.5$ & $2.5$ & $2.5$ \\
Bkgd suppression req.  & $0.8$ & $0.8$ & $3.0$ & $2.0$ & $3.0$ & $6.0$ \\
Mass fit model variation 	& $0.6$ & $0.8$ & $1.2$ & $0.9$ & $1.6$ & $2.6$ \\
Short-lifetime variation 					& $0.1$ & $0.2$ & $0.3$ & $0.2$ & $0.7$ & $1.7$ \\
Long-lifetime variation 					& $0.6$ & $1.0$ & $1.2$ & $0.3$ & $0.6$ & $0.9$ \\
Lifetime resolution model & $0.4$ & $1.5$ & $4.0$ & $0.6$ & $2.6$ & $3.4$ \\
\hline
Total systematic					& $3.5$ & $3.6$ & $6.4$ & $4.1$ & $4.9$ & $7.5$ \\
\hline
\hline
(2L-fit $-$ 1L-fit) / 2L-fit (prompt)		     & $-0.1$ & $-0.4$ & $-0.6$ & $-0.3$ & $-0.5$ & $-3.4$ \\
(2L-fit $-$ 1L-fit) / 2L-fit (non-prompt) & $+0.1$ & $+0.4$ & $+0.7$ & $+0.1$ & $+1.4$ & $+9.8$ \\
\hline
\end{tabular}
\caption{\label{tab:sysErrXS} Summary of relative uncertainties for the $\psi(2S)$ and $X(3872)$ cross-section measurements showing the smallest (Min), median (Med) and largest (Max) values across the $\pt$ bins. The last two rows are described in the text. The uncertainty of the integrated luminosity ($1.9\%$) is not included.
}
\end{center}
\end{table}

\begin{table}[h]
\renewcommand{\arraystretch}{1.1}
\begin{center}
\begin{tabular}{l||rrr|rrr|rrr}
\hline
& \multicolumn{9}{c}{Absolute uncertainty $[\%]$} \\
& \multicolumn{3}{c}{$f^{\psi}_{\mathrm{NP}}$ } &  \multicolumn{3}{c}{$f^{X}_{\mathrm{NP}}$}   &  \multicolumn{3}{c}{$f^{X}_{\mathrm{SL}}$ }\\
Source of uncertainty 						& Min & Med & Max & Min & Med & Max  & Min & Med & Max \\
\hline \hline
Statistical									    	& $0.4$ & $0.5$ & $1.4$ & $4.2$ & $5.8$ & $17.8$  & $16.4$ & $25.8$ & $63$\hspace{1.5mm} \\
\hline
Trigger eff. 					& $0.1$ & $0.1$ & $0.3$ & $0.1$ & $0.1$ & $0.4$ & $0.0$ & $0.1$ & $0.1$ \\
Muon tracking eff.	& $0.0$ & $0.0$ & $0.0$ & $0.0$ & $0.0$ & $0.0$ &$0.0$ & $0.0$ & $0.0$\\
Muon reconstruction eff. 		& $0.0$ & $0.0$ & $0.1$ & $0.0$ & $0.0$ & $0.1$ & $0.0$ & $0.0$ & $0.1$ \\
Pion reconstruction eff.	    	& $0.4$ & $0.5$ & $0.7$ & $0.3$ & $0.3$ & $0.4$ & $0.0$ & $0.3$ & $0.4$\\
Bkgd suppression req. & $0.8$ & $1.1$ & $1.4$ & $0.6$ & $0.7$ & $0.7$ & $0.1$ & $0.1$ & $0.7$  \\
Mass fit model variation 					& $0.1$ & $0.1$ & $0.2$ & $0.2$ & $0.6$ & $1.8$ & $1.0$ & $1.3$ & $2.4$ \\
Lifetime resolution variation			& $0.2$ & $0.7$ & $1.7$ & $0.4$ & $1.0$ & $2.9$ & $1.8$ & $3.6$ & $12.1$ \\
Short-lifetime variation 					& $0.0$ & $0.1$ & $0.1$ & $0.1$ & $0.4$ & $0.8$ & $0.3$ & $0.7$ & $2.8$ \\
Long-lifetime variation 					& $0.3$ & $0.4$ & $0.4$ & $0.2$ & $0.2$ & $0.3$ & $3.3$ & $4.0$ & $4.4$ \\ \hline
Total systematic								& $1.3$ & $1.5$ & $2.4$ & $1.0$ & $1.4$ & $3.6$ & $4.1$ &  $4.9$ & $13.5$  \\
\hline
\hline
(2L-fit $-$ 1L-fit) / 2L-fit		   						& $+0.4$ & $+0.6$ & $+0.9$ & $+0.9$ & $+3.1$ & $+9.1$ & $-$ & $-$ & $-$ \\
\hline
\end{tabular}
\caption{\label{tab:sysErrFrac}  Summary of uncertainties for $\psi(2S)$ and $X(3872)$ non-prompt fractions, and short-lived non-prompt fraction for $X(3872)$ production, showing the smallest (Min), median (Med) and largest (Max) values across the $\pt$ bins. The last row is described in the text.}
\end{center}
\end{table}

Uncertainties in the trigger efficiency, and in the muon and pion reconstruction efficiencies are determined using the procedures adopted in~Ref.~\cite{BPHY-2013-06}.
Additional uncertainty of $\pm 2 \%$~\cite{BPHY-2012-02b} is assigned to the tracking efficiency of the two muons within the ID, primarily due to its dependence on the total number of $pp$ collisions per event.
The uncertainties in matching generator-level particles to reconstruction-level particles, and in the detector material simulation within the barrel part of the inner detector are found to be the main contributions to the systematic uncertainty of the pion reconstruction efficiency, estimated to be
$\pm 2.5 \%$.
Such efficiency uncertainties largely cancel in the various non-prompt fractions (Table~\ref{tab:sysErrFrac}).

The uncertainties in the efficiency of the background suppression requirements (see Section~\ref{sec:crossSection}), obtained by combining MC statistical errors and systematic errors in quadrature, are in the range $1\%$--$6\%$.
The uncertainties in the mass fits are estimated by varying the values of parameters that were fixed during the main fit, and by increasing the order of the polynomial $P$ in the background parameterisation (see Equation~(\ref{mainFitEqn})).
Similarly, the systematic uncertainties of the lifetime fits are determined by varying the
values of the fixed lifetimes and the parameters of the lifetime resolution function within their predetermined ranges.

The statistical and individual systematic uncertainties are added in quadrature to form the total error shown in the tables.
In general, the results for $X(3872)$ are dominated by statistical errors, while  for $\psi(2S)$ statistical and systematic uncertainties are of comparable size.

The last rows in Tables \ref{tab:sysErrXS} and \ref{tab:sysErrFrac} show the relative differences between the values obtained using the single- and two-lifetime fits, labelled as `1L-fit' and `2L-fit', respectively. For the quantities listed in Tables \ref{tab:sysErrXS} and \ref{tab:sysErrFrac}, these differences were found to be generally fairly small, compared to the combined systematic uncertainty from other sources.

\section{Results and discussion}
\label{sec:result}

\begin{figure}[h]
\begin{center}
\subfigure[]{\includegraphics[width=0.48\textwidth]{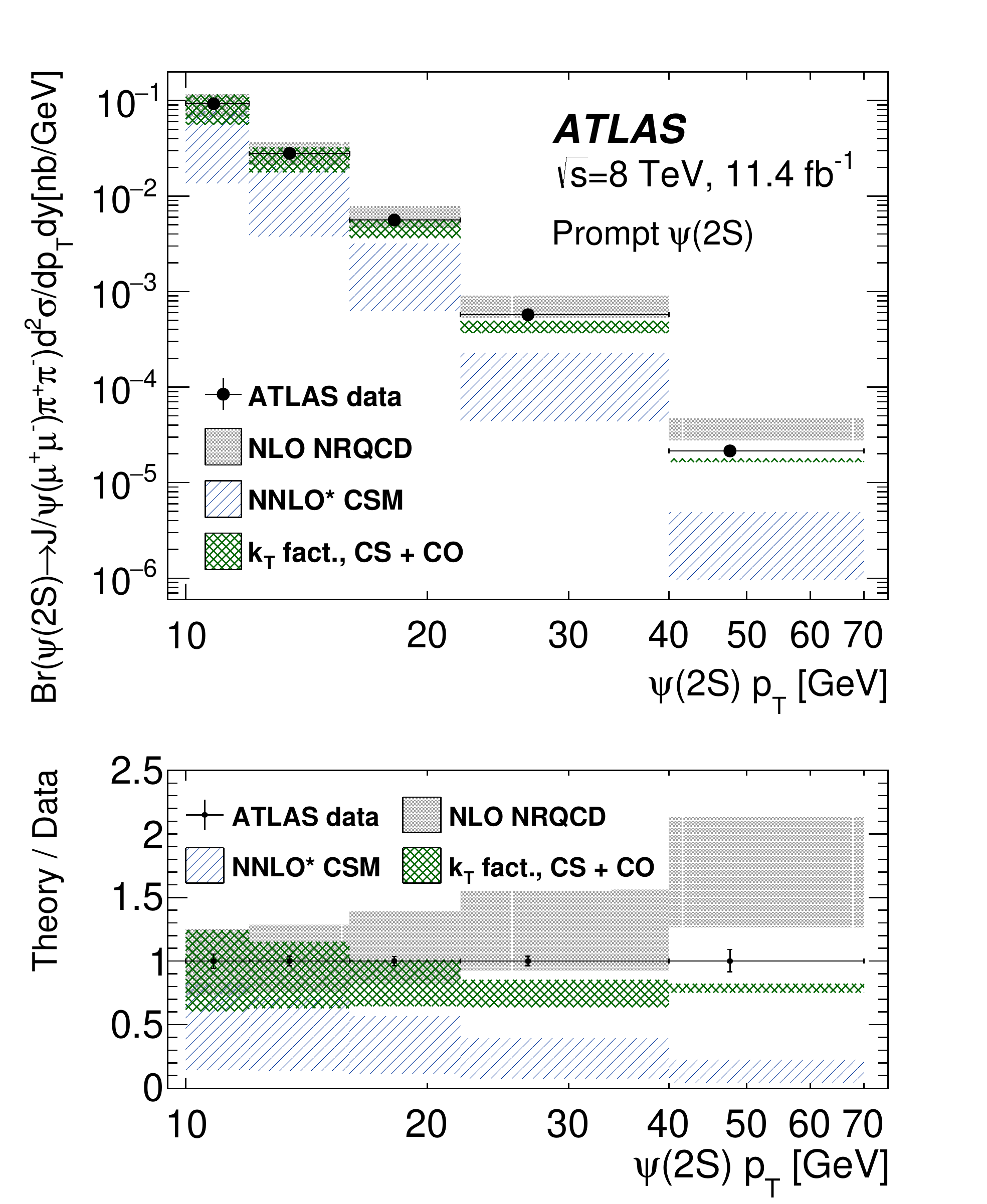}\label{prompt_psi2S_XS}}
\subfigure[]{\includegraphics[width=0.48\textwidth]
{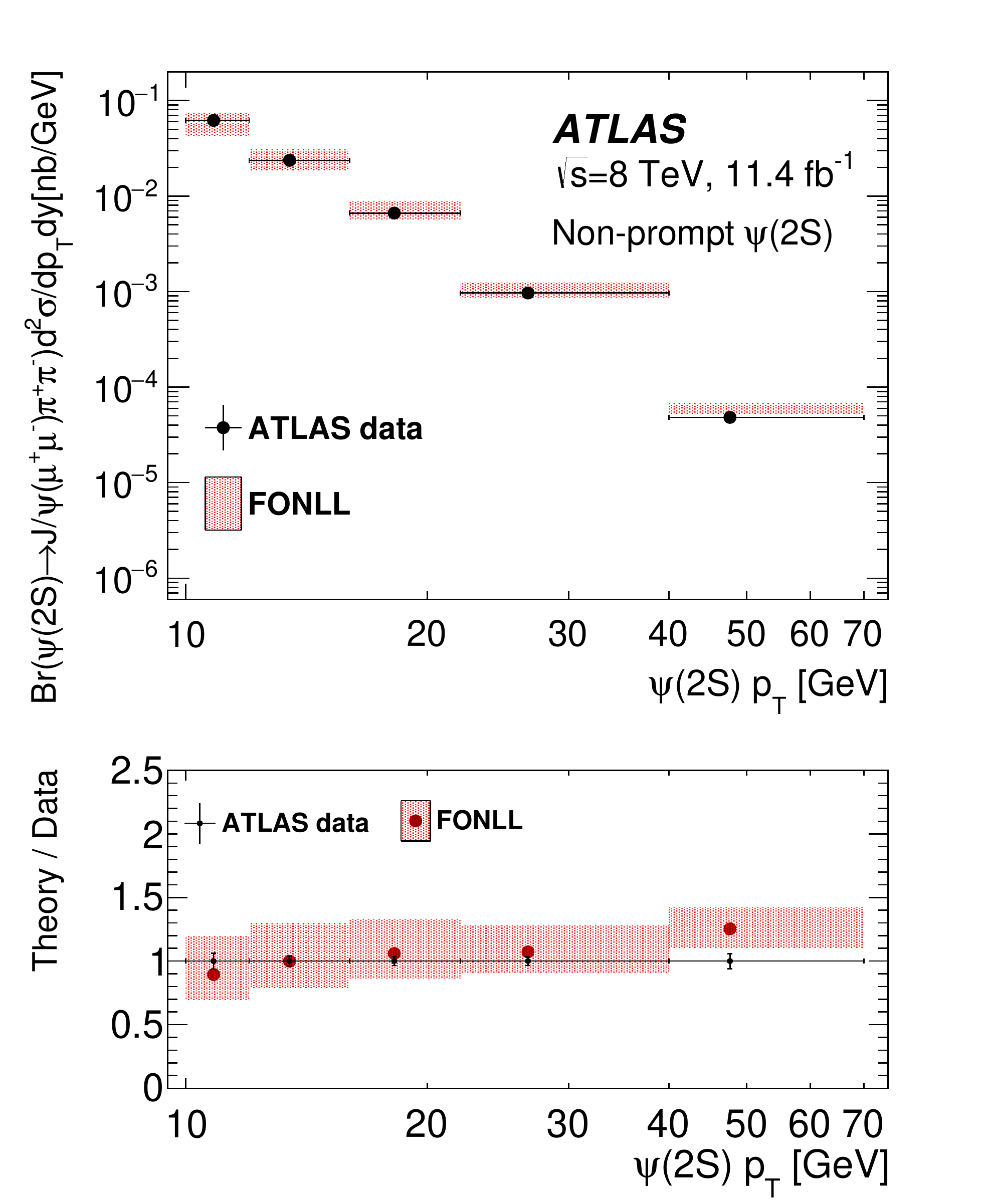}\label{nonprompt_psi2S_XS}}
\caption{Measured cross section times branching fractions as a function of $\pt$ for \subref{prompt_psi2S_XS} prompt $\psi(2S)$ production compared to NLO NRQCD~\cite{BPHY-2013-01}, the $k_{\mathrm{T}}$ factorisation model \cite{Baranov:2015laa} and the NNLO* CSM~\cite{Lansberg:2008gk}, and \subref{nonprompt_psi2S_XS} non-prompt $\psi(2S)$ production compared to FONLL~\cite{Cacciari:2012ny} predictions.}
\label{psi2S_XS}
\end{center}
\end{figure}

The measured differential cross section (times the product of the relevant branching fractions)
for prompt production of $\psi(2S)$ is shown in Figure \ref{prompt_psi2S_XS}. It is described fairly well by the NLO NRQCD model~\cite{BPHY-2013-01} with long-distance matrix elements (LDMEs) determined from the Tevatron data, although some overestimation is observed at the highest $\pt$ values.
The $k_{\mathrm{T}}$ factorisation model \cite{Baranov:2015laa}, which includes the colour-octet (CO) contributions tuned to $7 \TeV$ CMS data~\cite{Khachatryan:2015rra} in addition to colour-singlet (CS) production, describes ATLAS data fairly well, with a slight underestimation at higher $\pt$.
The NNLO* Colour-Singlet Model (CSM) predictions~\cite{Lansberg:2008gk} are close to the data points at low $\pt$, but significantly underestimate them at higher $\pt$ values. The measured differential cross section  for non-prompt $\psi(2S)$ production is presented in Figure \ref{nonprompt_psi2S_XS}, compared with the predictions of
the FONLL calculation~\cite{Cacciari:2012ny}. The calculation describes the data well over the whole range of transverse momenta.

\begin{figure}[h]
\begin{center}
\subfigure[]{\includegraphics[width=0.48\textwidth]{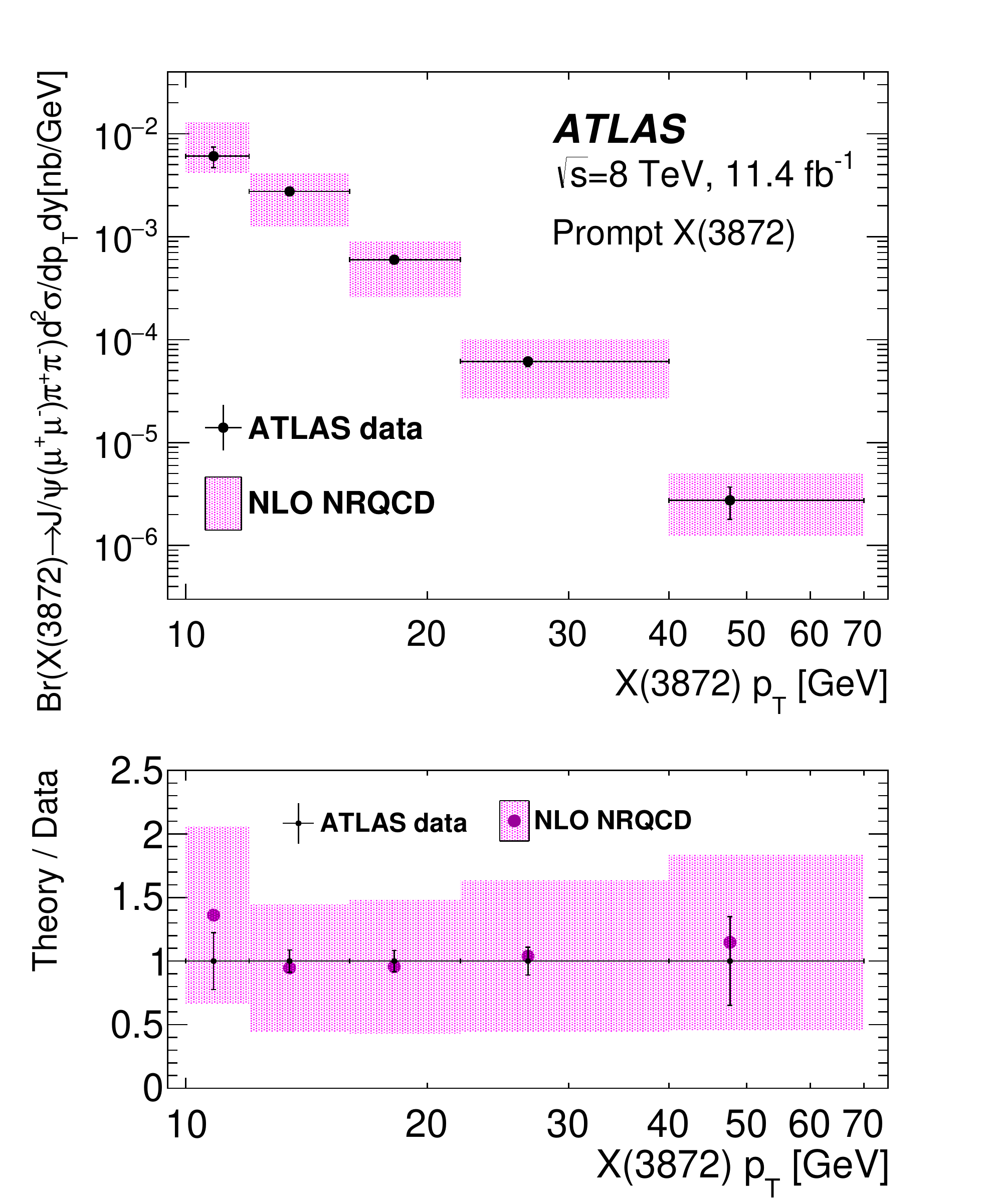}\label{prompt_X3872_XS}}
\subfigure[]{\includegraphics[width=0.48\textwidth]{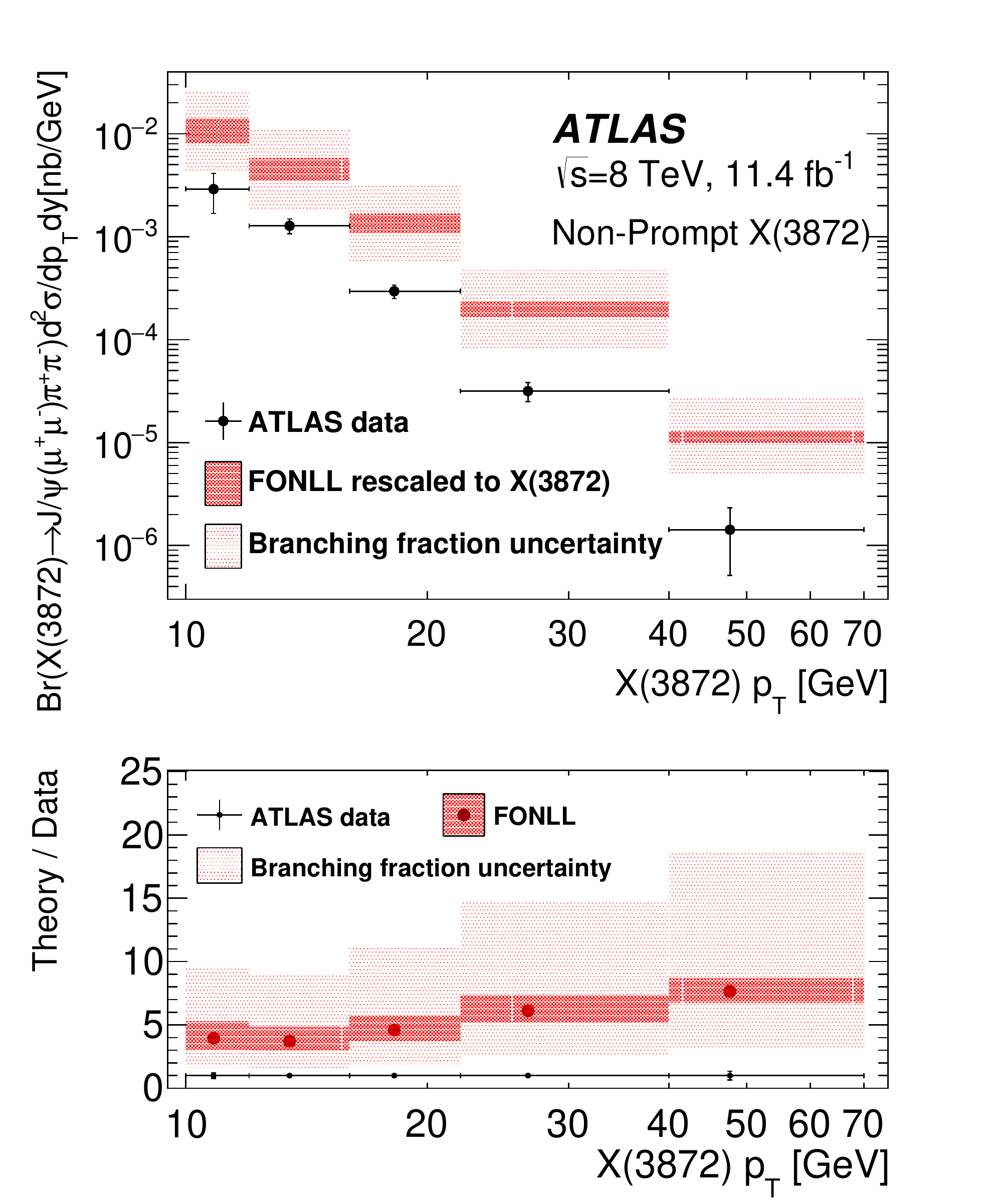}\label{nonprompt_X3872_XS}}
\caption{Measured cross section times branching fractions as a function of $\pt$ for \subref{prompt_X3872_XS} prompt $X(3872)$ compared to NLO NRQCD predictions with the $X(3872)$ modelled as a mixture of $\chi_{c1}(2P)$ and a $D^{0}\bar{D}^{*0}$ molecular state~\cite{chidd}, and \subref{nonprompt_X3872_XS} non-prompt $X(3872)$ compared to the FONLL~\cite{Cacciari:2012ny} model prediction, recalculated using the branching fraction estimate from~Ref.~\cite{Artoisenet:2009wk} as described in the text.}
\label{X3872_XS}
\end{center}
\end{figure}

Similarly, the differential cross section
for prompt production of $X(3872)$ is shown in
Figure~\ref{prompt_X3872_XS}. It is described within the theoretical uncertainty by the prediction of the NRQCD model which, in this case, considers $X(3872)$ to be a mixture of $\chi_{c1}(2P)$ and a  $D^{0}\bar{D}^{*0}$ molecular state~\cite{chidd}, with the production being dominated by the
$\chi_{c1}(2P)$ component and the normalisation fixed through the fit to CMS
data~\cite{cmsMeas}. The measured differential
cross section  for non-prompt production of $X(3872)$ is
shown in Figure~\ref{nonprompt_X3872_XS}.
This is compared to a calculation based on the FONLL model prediction for $\psi(2S)$, recalculated for
$X(3872)$
using the kinematic template for the non-prompt $X(3872) / \psi(2S)$ ratio shown in
Figure~\ref{fig:npr} and
the effective value of the product of the branching fractions
$\mathcal{B}(B \to X(3872)) \mathcal{B}(X(3872) \to \jpsi\pi^+\pi^-)=(1.9\pm0.8) \times 10^{-4}$
estimated in~Ref.~\cite{Artoisenet:2009wk} based on the Tevatron data~\cite{Bauer:2004bc}.
This calculation overestimates the data by a factor increasing with $\pt$ from about four to about eight over the $\pt$ range of this measurement.

\begin{figure}[H]
\begin{center}
\subfigure[]{ \includegraphics[width=0.47\textwidth]{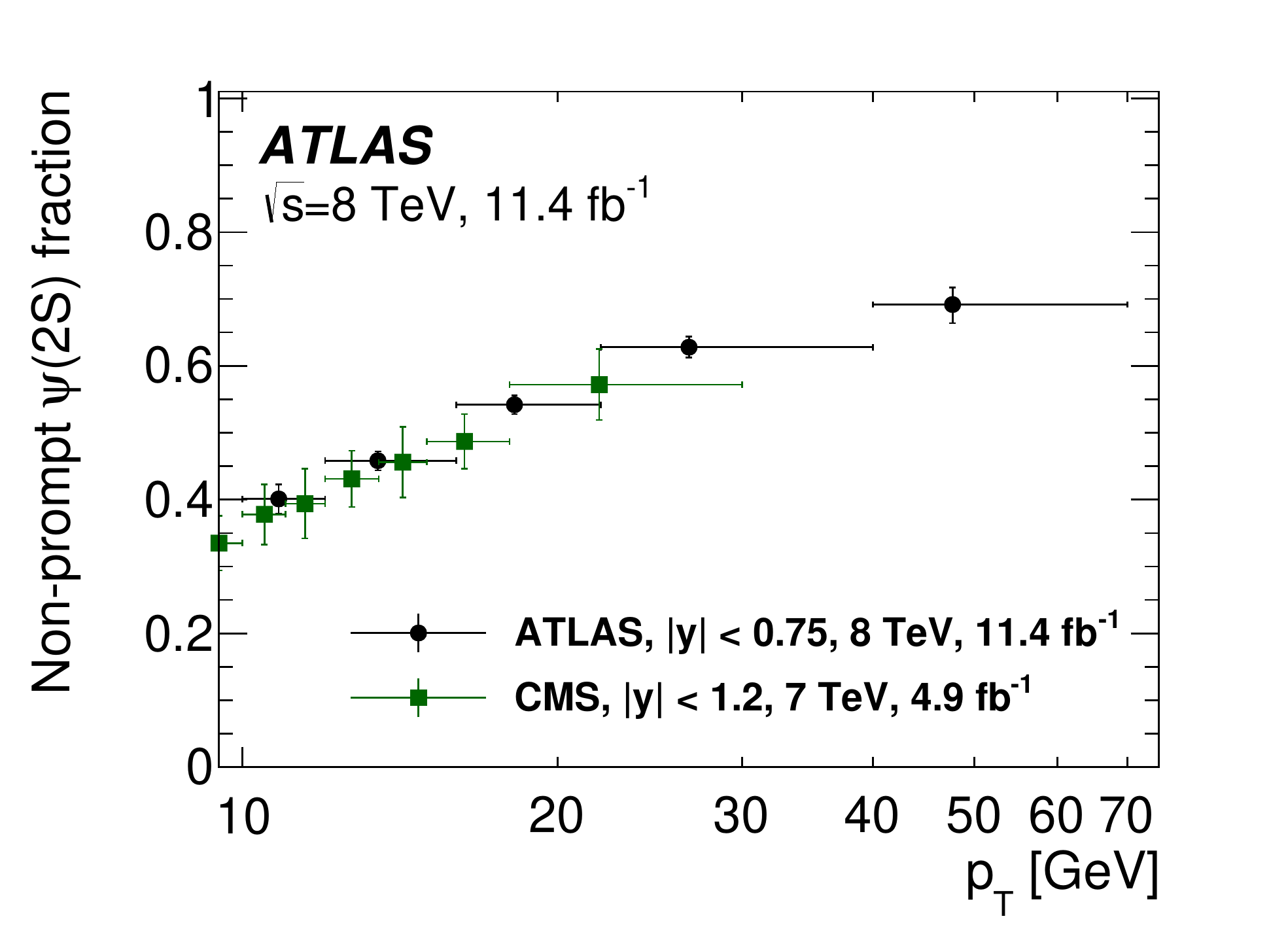}\label{compnpfPsi2SXSexperimentSLTL} }
\subfigure[]{ \includegraphics[width=0.47\textwidth]{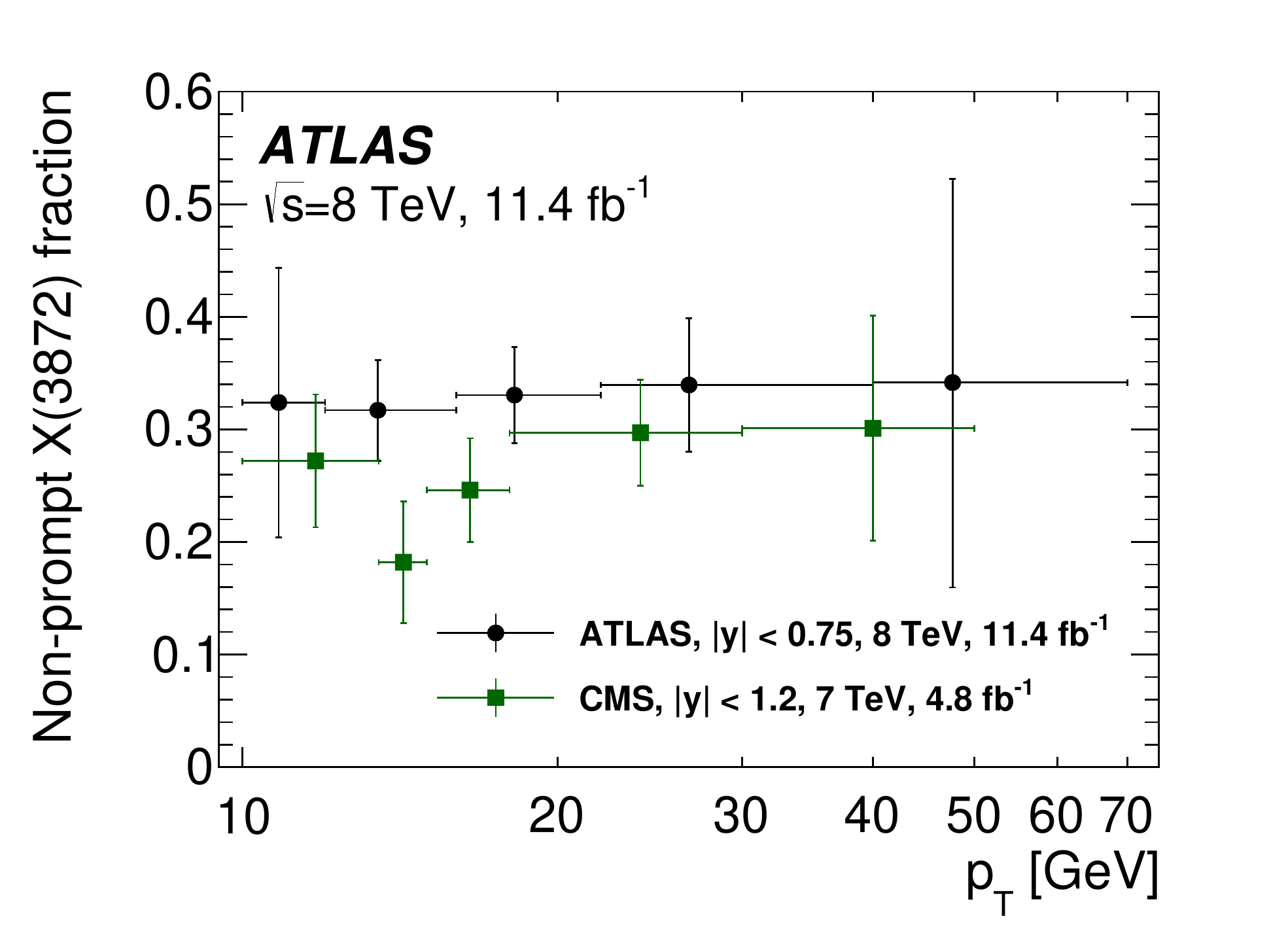}\label{compnpfX3872XSexperimentSLTL} }
\caption{Measured non-prompt fractions for \subref{compnpfPsi2SXSexperimentSLTL} $\psi(2S)$ and \subref{compnpfX3872XSexperimentSLTL} $X(3872)$ production, compared to CMS results at $\sqrt{s} = 7$~TeV. The blue circles are the results shown in this paper, while the green squares show CMS results~\cite{cmsPsi2S2011,cmsMeas}.}
\label{compnpfXSexperimentSLTL}
\end{center}
\end{figure}

The non-prompt fractions of $\psi(2S)$ and $X(3872)$ production are shown in Figure \ref{compnpfXSexperimentSLTL}. In the case of $\psi(2S)$, $f_{\mathrm{NP}}$ increases with $\pt$, in good agreement with measurements obtained with dimuon decays of $\psi(2S)$ from ATLAS~\cite{BPHY-2012-02b} and CMS~\cite{cmsPsi2S2011}. The non-prompt fraction of $X(3872)$ shows no sizeable dependence on $\pt$. This measurement agrees within errors with the CMS result obtained at $\sqrt{s} = $7~TeV~\cite{cmsMeas}. 

The numerical values of all cross sections and fractions shown in Figures \ref{fig:ratios}--\ref{compnpfXSexperimentSLTL} are presented in Table~\ref{tab:XS_ratios_TL}.

\begin{sidewaystable}[!htbp]
\renewcommand{\arraystretch}{1.4}
\footnotesize
\begin{center}
\begin{tabular}{c r@{\hspace{1mm}$\pm$\hspace{1mm}}l@{\hspace{1mm}$\pm$\hspace{1mm}}l r@{\hspace{1mm}$\pm$\hspace{1mm}}l@{\hspace{1mm}$\pm$\hspace{1mm}}l r@{\hspace{1mm}$\pm$\hspace{1mm}}l@{\hspace{1mm}$\pm$\hspace{1mm}}l r@{\hspace{1mm}$\pm$\hspace{1mm}}l@{\hspace{1mm}$\pm$\hspace{1mm}}l r@{\hspace{1mm}$\pm$\hspace{1mm}}l@{\hspace{1mm}$\pm$\hspace{1mm}}l}
\hline
& \multicolumn{15}{c}{$p_{\rm T}$ range [GeV]} \\
&\multicolumn{3}{c}{10--12} &  \multicolumn{3}{c}{12--16}  &  \multicolumn{3}{c}{16--22}  &  \multicolumn{3}{c}{22--40} &  \multicolumn{3}{c}{40--70}  \\  \hline
& \multicolumn{15}{c}{Cross sections times branching fractions [pb / GeV]} \\ \hline
$\psi(2S)_{\mathrm{P}}$ & $ 92.4$ & $1.9$ & $ 4.8$&$27.97$ & $0.27$ & $ 1.02$&$5.61$ & $0.06$ & $ 0.19$&$0.57$ & $0.01$ & $ 0.02$&$0.021$ & $0.001$ & $ 0.001$ \\ 
$\psi(2S)_{\mathrm{NP}}$ & $ 61.9$ & $1.9$ & $ 3.4$&$23.66$ & $0.27$ & $ 0.85$&$6.63$ & $0.06$ & $ 0.22$&$0.97$ & $0.01$ & $ 0.03$&$0.048$ & $0.001$ & $ 0.003$ \\ 
$\psi(2S)^{\mathrm{LL}}_{\mathrm{NP}}$  & $ 60.8$ & $1.6$ & $ 4.0$&$23.09$ & $0.27$ & $ 1.46$&$6.53$ & $0.06$ & $ 0.41$&$0.93$ & $0.01$ & $ 0.06$&$0.047$ & $0.002$ & $ 0.003$ \\ 
$\psi(2S)^{\mathrm{SL}}_{\mathrm{NP}}$ & $ 1.1$ & $2.4$ & $ 3.9$&$0.56$ & $0.37$ & $ 1.14$&$0.11$ & $0.08$ & $ 0.29$&$0.04$ & $0.01$ & $ 0.04$&$0.001$ & $0.002$ & $ 0.002$ \\ 
$X(3872)_{\mathrm{P}}$ & $ 6.05$ & $1.30$ & $ 0.38$&$2.75$ & $0.20$ & $ 0.13$&$0.60$ & $0.04$ & $ 0.02$&$0.06$ & $0.01$ & $ 0.00$&$0.003$ & $0.001$ & $ 0.000$ \\ 
$X(3872)_{\mathrm{NP}}$ & $ 2.90$ & $1.20$ & $ 0.21$&$1.28$ & $0.20$ & $ 0.07$&$0.29$ & $0.04$ & $ 0.01$&$0.03$ & $0.01$ & $ 0.00$&$0.001$ & $0.001$ & $ 0.000$ \\ 
$X(3872)^{\mathrm{LL}}_{\mathrm{NP}}$ & $ 1.87$ & $0.82$ & $ 0.14$&$0.92$ & $0.16$ & $ 0.06$&$0.29$ & $0.04$ & $ 0.02$&$0.03$ & $0.01$ & $ 0.00$&$0.001$ & $0.001$ & $ 0.000$ \\ 
$X(3872)^{\mathrm{SL}}_{\mathrm{NP}}$ & $ 1.02$ & $1.49$ & $ 0.20$&$0.35$ & $0.25$ & $ 0.06$&$0.01$ & $0.06$ & $ 0.02$&$0.00$ & $0.01$ & $ 0.00$&$0.000$ & $0.001$ & $ 0.000$ \\   \hline
& \multicolumn{15}{c}{Fractions} \\ \hline
$F^{\psi(2S)}_{\mathrm{NP}}$ & $ 0.40$ & $0.01$ & $ 0.02$&$0.46$ & $0.00$ & $ 0.01$&$0.54$ & $0.00$ & $ 0.01$&$0.63$ & $0.00$ & $ 0.01$&$0.69$ & $0.01$ & $ 0.02$ \\ 
$F^{\psi(2S)}_{\mathrm{SL}}$ & $ 0.02$ & $0.04$ & $ 0.06$&$0.02$ & $0.02$ & $ 0.05$&$0.02$ & $0.01$ & $ 0.04$&$0.04$ & $0.01$ & $ 0.04$&$0.03$ & $0.03$ & $ 0.05$ \\ 
$F^{X(3872)}_{\mathrm{NP}}$ & $ 0.32$ & $0.12$ & $ 0.02$&$0.32$ & $0.04$ & $ 0.01$&$0.33$ & $0.04$ & $ 0.01$&$0.34$ & $0.06$ & $ 0.01$&$0.34$ & $0.18$ & $ 0.03$ \\ 
$F^{X(3872)}_{\mathrm{SL}}$ & $ 0.35$ & $0.39$ & $ 0.05$&$0.28$ & $0.16$ & $ 0.04$&$0.03$ & $0.19$ & $ 0.05$&$0.03$ & $0.26$ & $ 0.05$&$0.03$ & $0.63$ & $ 0.13$ \\   \hline
& \multicolumn{15}{c}{Ratios} \\ \hline
$X(3872)_{\mathrm{P}} / \psi(2S)_{\mathrm{P}}$ & $ 0.065$ & $0.014$ & $ 0.004$&$0.098$ & $0.007$ & $ 0.004$&$0.106$ & $0.008$ & $ 0.004$&$0.107$ & $0.011$ & $ 0.004$&$0.128$ & $0.044$ & $ 0.012$ \\ 
$X(3872)_{\mathrm{NP}} / \psi(2S)_{\mathrm{NP}}$ & $ 0.047$ & $0.019$ & $ 0.004$&$0.054$ & $0.008$ & $ 0.003$&$0.044$ & $0.006$ & $ 0.002$&$0.033$ & $0.007$ & $ 0.001$&$0.030$ & $0.019$ & $ 0.003$ \\ 
$X(3872)_{\mathrm{NP}}^{\mathrm{LL}} / \psi(2S)_{\mathrm{NP}}^{\mathrm{LL}}$ & $ 0.031$ & $0.014$ & $ 0.002$&$0.040$ & $0.007$ & $ 0.003$&$0.044$ & $0.006$ & $ 0.003$&$0.033$ & $0.006$ & $ 0.002$&$0.030$ & $0.019$ & $ 0.003$ \\ 
$X(3872)_{\mathrm{NP}}^{\mathrm{SL}} / \psi(2S)_{\mathrm{NP}}^{\mathrm{LL}}$ & $ 0.016$ & $0.024$ & $ 0.003$&$0.015$ & $0.011$ & $ 0.003$&$0.001$ & $0.008$ & $ 0.002$&$0.001$ & $0.009$ & $ 0.004$&$0.001$ & $0.024$ & $ 0.005$ \\ 
\hline
\end{tabular}
\end{center}
\caption{\label{tab:XS_ratios_TL} Summary of $\psi(2\mathrm{S})$ and $X(3872)$ cross-section measurements, fractions and ratios.
The subscripts P and NP denote prompt and non-prompt components, while the labels SL and LL stand for short-lived and long-lived non-prompt components, respectively.
The first uncertainty is statistical, the second is systematic. Uncertainties from integrated luminosity ($1.9\%$) and those due to unknown spin-alignment are not included.}
\end{sidewaystable}

\section{Dipion invariant mass spectra}
\label{sec:dipionIntervals2}

The distributions of the dipion invariant mass $m_{\pi\pi}$ in the $\psi(2S)\rightarrow J/\psi\pi^+\pi^-$ and $X(3872) \rightarrow J/\psi\pi^+\pi^-$ decays are measured by determining the corrected yields of $\psi(2S)$ and $X(3872)$  signals in narrow bins of $m_{\pi\pi}$. The two additional selection requirements (Equation~(\ref{backgroundCuts})) used specifically to reduce
combinatorial background in the cross-section measurement, are found to bias the $m_{\pi\pi}$ distributions and are therefore replaced for this study by requirements
on the pseudo-proper lifetime significance, $\tau/\Delta \tau < 2.5$,
and the transverse momentum
of the $J/\psi \pi^+ \pi^-$ candidates, $\pt > 12$~GeV.

\begin{figure}[h]
\begin{center}
\subfigure[]{\includegraphics[width=0.49\textwidth]{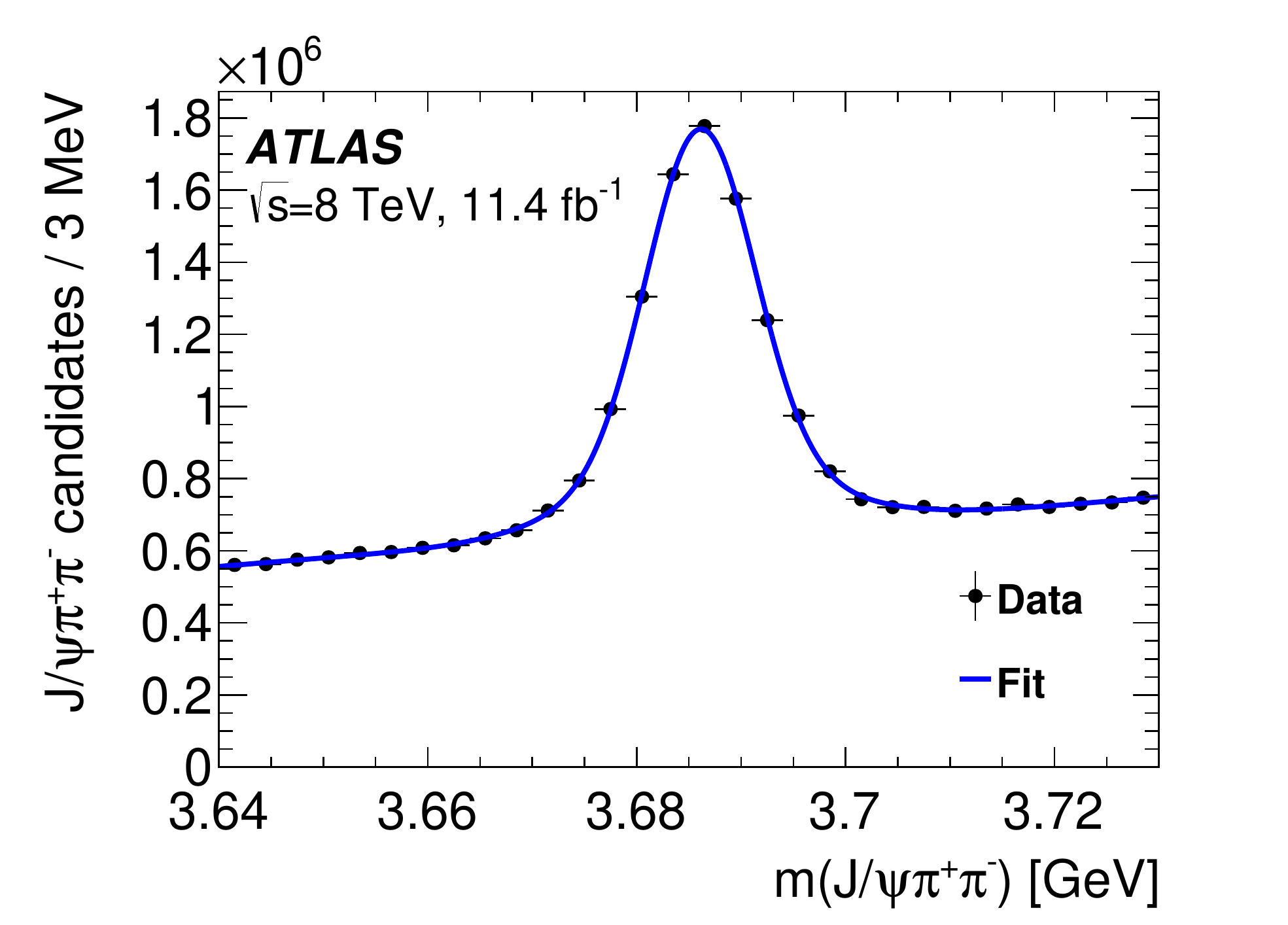}\label{fig:invMassPsi2Smpipi} }
\subfigure[]{\includegraphics[width=0.49\textwidth]{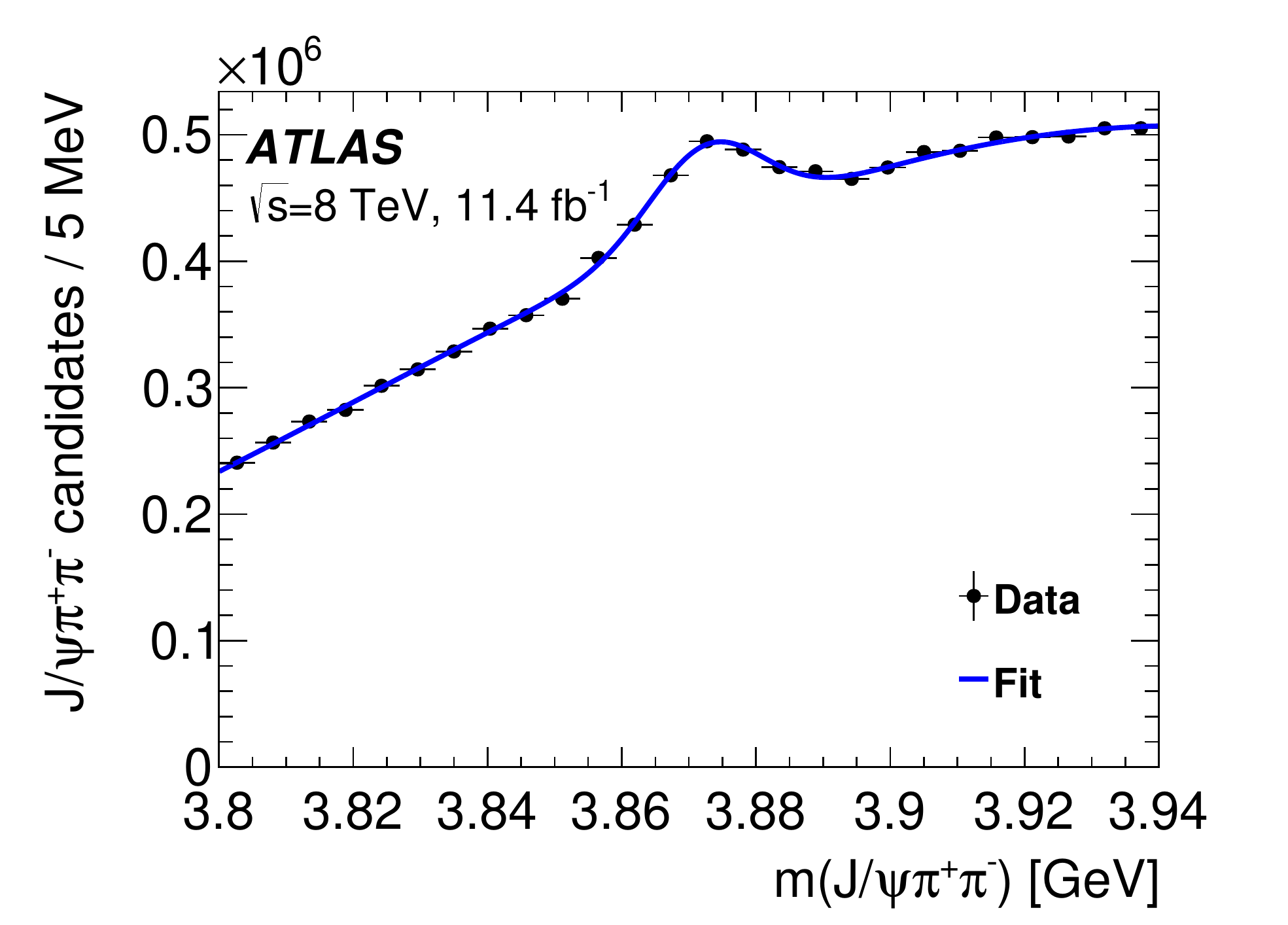}\label{fig:invMassX3872mpipi} }
\caption{The invariant mass distributions of the $\jpsi\pi^+\pi^-$ candidates to extract \subref{fig:invMassPsi2Smpipi} $\psi(2S)$ and \subref{fig:invMassX3872mpipi} $X(3872)$ signal integrated over  a wide range of $m_{\pi\pi}$.}
\label{fig:invMassmpipi}
\end{center}
\end{figure}

The invariant mass distributions of the corrected $J/\psi\pi^+\pi^-$ candidates selected for this analysis are shown in Figure~\ref{fig:invMassPsi2Smpipi} for the mass range around $\psi(2S)$ peak and in Figure~\ref{fig:invMassX3872mpipi} for $X(3872)$.

The interval of allowed $m_{\pi\pi}$
values is subdivided into 21 and 11 bins for $\psi(2S)$ and $X(3872)$, respectively.
In each $m_{\pi\pi}$ bin, the signal yield is extracted using a fit to the function
\begin{equation}
f(m) = Y\left[ f_1G_{1}(m) + (1-f_1)G_{2}(m)\right]+ N_{\mathrm{bkg}}\left(  \frac{m - p_0}{m_{0}  - p_0} \right)^{p_1}\mathrm{e}^{-p_2(m -p_0)-p_3(m - p_0)^2 } ,
\label{dipionFittingFcn}
\end{equation}
where $m$ is the invariant mass of the $J/\psi\pi^+\pi^-$ system, $Y$ is the yield of the parent resonance, $N_{\mathrm{bkg}}$ is the normalisation factor of the background PDF,
$m_{0}$ is the world average mass~\cite{PDGBcTau} of the parent resonance, and
$p_{0,1,2,3}$ are free parameters.
The signals are described by the same double-Gaussian PDFs
$ f_1G_{1}(m) + (1-f_1)G_{2}(m)$ as the ones used in the cross-section analysis
described in Section \ref{sec:crossSection}.
In most $m_{\pi\pi}$ bins the position of the signal peak is determined from the fit; however, in some bins with small signal yields it is necessary to fix the centre and the width of the signal peak to the values obtained from the fits over the whole $m_{\pi\pi}$ range shown in Figure \ref{fig:invMassX3872mpipi}.
As in the cross-section analysis, the fraction of the
narrow Gaussian function $f_{1}$ is fixed to $0.76\pm0.04$,
varied within the range of $\pm0.04$ during systematic uncertainty studies.
In another variation a first-order polynomial is
added as a factor multiplying the PDF in Equation~(\ref{dipionFittingFcn}).
For both the $\psi(2S)$ and $X(3872)$ samples, the errors from the fits in $m_{\pi\pi}$ bins are found to be statistically dominated.

\begin{figure}
\begin{center}
\subfigure[]{\includegraphics[width=0.49\textwidth]{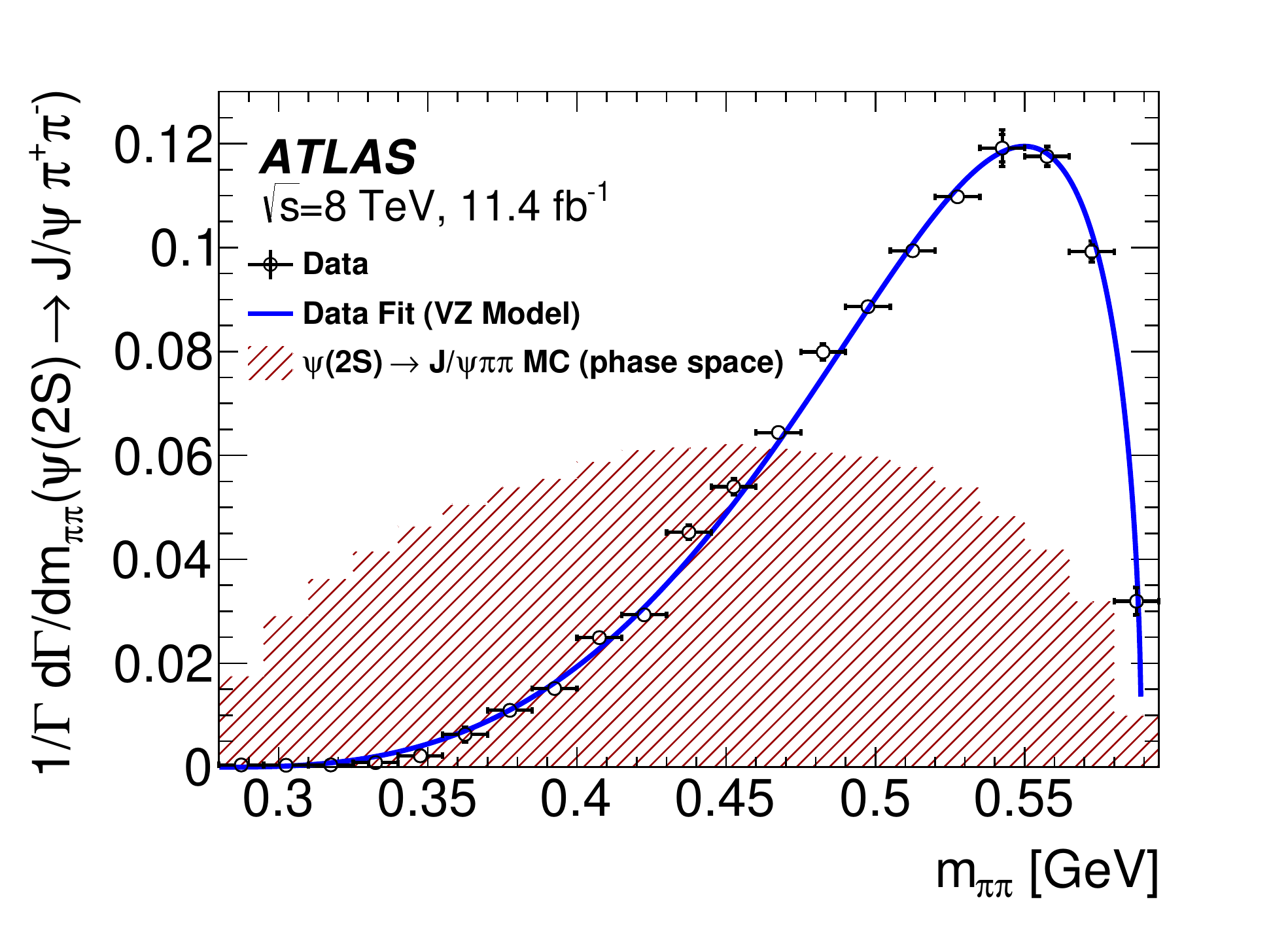}\label{dipionComparison_Psi2S} }
\subfigure[]{\includegraphics[width=0.49\textwidth]{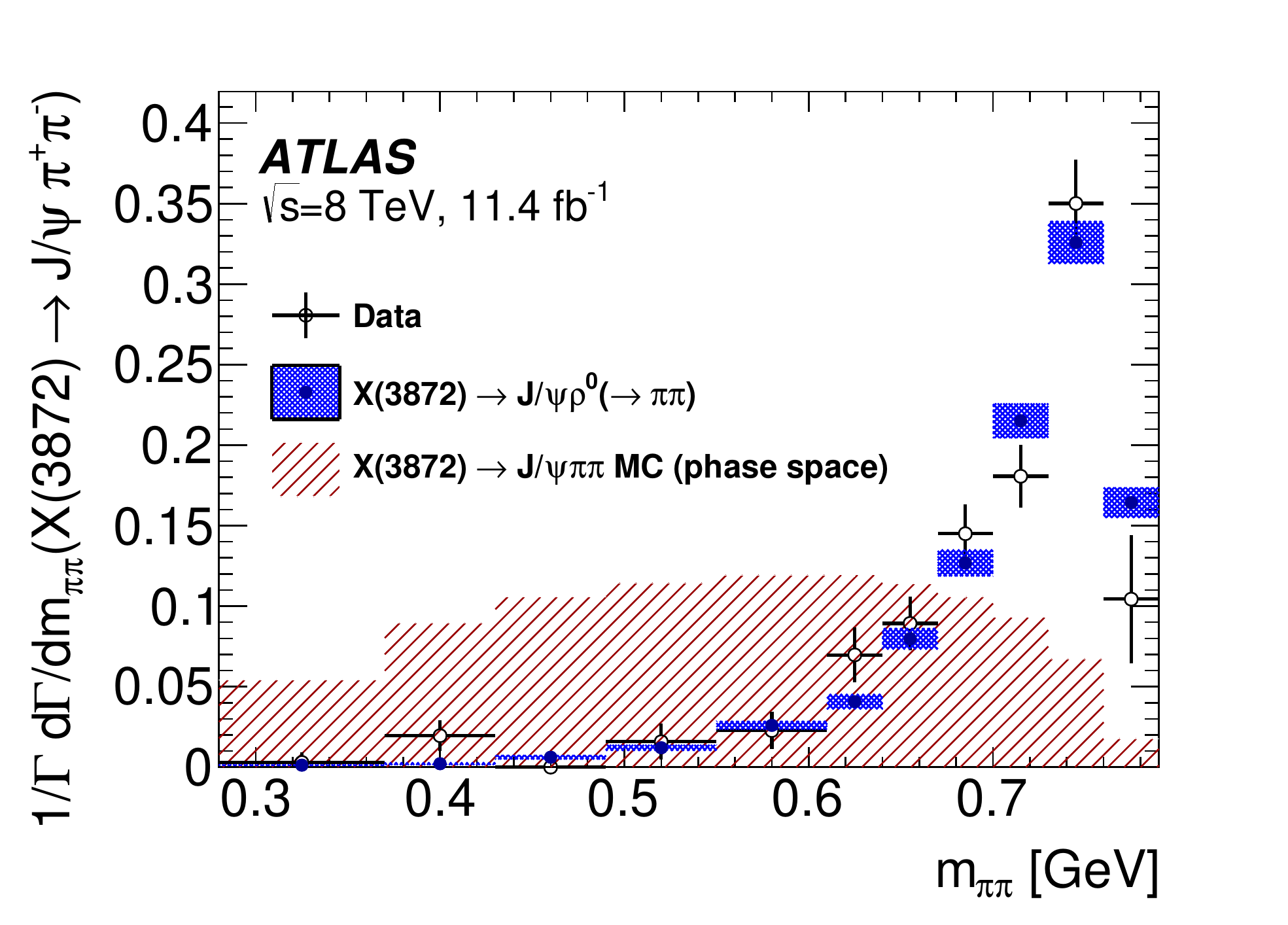}\label{dipionComparison_b} }
\caption{\subref{dipionComparison_Psi2S} Normalised differential decay width of $\psi(2S) \rightarrow J/\psi(\rightarrow\mu^+\mu^-)\pi^+\pi^-$ in bins of dipion invariant mass over the range $ 0.280 \GeV < m_{\pi\pi} < 0.595 \GeV$, fitted with the Voloshin--Zakharov model. Also shown is the normalised $m_{\pi\pi}$ phase-space distribution (red shaded histogram). \subref{dipionComparison_b} Normalised differential decay width of $X(3872) \rightarrow J/\psi(\rightarrow\mu^+\mu^-)\pi^+\pi^-$ in bins of dipion invariant mass over the range $ 0.28\GeV < m_{\pi\pi} < 0.79\GeV$. Also shown is the MC prediction for the decay $X(3872) \rightarrow J/\psi(\rightarrow\mu^+\mu^-)\rho^0(\rightarrow \pi^+\pi^-)$ (blue histogram)  and the normalised distribution of $m_{\pi\pi}$ phase-space (red shaded histogram).}
\label{fig:dipionComparison}
\end{center}
\end{figure}

The resulting normalised differential distributions in $m_{\pi\pi}$ are shown in Figure~\ref{dipionComparison_Psi2S} for $\psi(2S)\to J/\psi\pi^+\pi^-$ and in Figure~\ref{dipionComparison_b} for $X(3872)\to J/\psi\pi^+\pi^-$ decays.
 The solid blue curve in Figure~\ref{dipionComparison_Psi2S} represents a fit
to the data points with the Voloshin--Zakharov distribution~\cite{Voloshin:1980zf}
\begin{equation}
\frac{1}{\Gamma}\frac{\mathrm{d}\Gamma}{\mathrm{d}m_{\pi\pi}} \propto \left( m_{\pi\pi}^2 - \lambda m_{\pi}^2 \right)^2 \times {\mathrm{PS}},
\end{equation}
where ${\mathrm{PS}}$ stands for the dipion phase-space.
The fitted value of the parameter $\lambda$ is found to be $\lambda = 4.16 \pm 0.06\mathrm{(stat)} \pm 0.03\mathrm{(sys)}$, in agreement with $\lambda = 4.35 \pm 0.18$ measured by BES~\cite{voloshinBES}, and $\lambda = 4.46 \pm 0.25$ measured by LHCb~\cite{voloshinLHCb2}. The shaded blue histogram in Figure \ref{dipionComparison_b} is obtained from straightforward simulations, assuming the dipion system in the decay $X(3872)\to J/\psi\pi^+\pi^-$ is produced purely via the $\rho^0$ meson, and appears to be in good agreement with the data. In both decays the measured $m_{\pi\pi}$ spectrum strongly disfavours the dipion phase-space distribution (shown in Figures \ref{dipionComparison_Psi2S} and \ref{dipionComparison_b} by the red shaded area), with the data clearly preferring higher masses in either case.

\section{Summary}
\label{sec:conclusion}

The measurement of the differential production cross section of $\psi(2S)$ and $X(3872)$ states
in the $\jpsi\pi^+\pi^-$ final state
is carried out using 11.4~fb$^{-1}$ of $\sqrt{s}= 8$~TeV $pp$ collision data recorded by the ATLAS detector at the LHC.
The prompt and non-prompt production of $\psi(2S)$ and $X(3872)$
is studied separately, as a function of transverse momentum in the rapidity region $|y| < 0.75$ and
transverse momentum range $10 \GeV < \pt < 70 \GeV$.

The $\psi(2S)$ cross-section measurements show good consistency with the theoretical
predictions based on NLO NRQCD and FONLL for prompt and non-prompt production, respectively. The predictions from the $k_{\mathrm{T}}$ factorisation model with the colour-octet component tuned to 7~TeV CMS data describe the prompt $\psi(2S)$ measurement fairly well, while NNLO* colour-singlet model calculations underestimate the data, especially at higher transverse momenta.

The prompt $X(3872)$ cross-section measurement
shows good agreement with the CMS result for transverse momenta $10\GeV < \pt < 30\GeV$
where they overlap, and extends the range of transverse momenta up to $70\GeV$.
Good agreement is found with theoretical predictions within the model based on NLO NRQCD, which considers $X(3872)$ to be a mixture of $\chi_{c1}(2P)$ and a $D^{0}\bar{D}^{*0}$ molecular state, with the production being dominated by the $\chi_{c1}(2P)$ component and the normalisation fixed through the fit to CMS data.

The non-prompt production of $\psi(2S)$ is described by the FONLL predictions within the uncertainties.
But the same predictions, recalculated for $X(3872)$  using the branching fraction extracted from the Tevatron data, overestimate the non-prompt production of $X(3872)$, especially at large transverse momenta.

Two models of lifetime dependence of the non-prompt production are considered: a model with a single effective lifetime, and an alternative model with two distinctly different effective lifetimes. The two models give compatible results for the prompt and non-prompt differential cross sections of $\psi(2S)$ and $X(3872)$.

Within the single-lifetime model, assuming that non-prompt $\psi(2S)$ and $X(3872)$ originate from the same mix of parent $b$-hadrons, the following result is obtained for the ratio of the branching fractions:
\begin{equation}
R^{\mathrm{1L}}_{B} =  \frac{\mathcal{B}(B \rightarrow X(3872) \textrm{ + any})  \mathcal{B}(X(3872) \rightarrow \jpsi\pi^+\pi^-)}{\mathcal{B}(B \rightarrow \psi(2S) \textrm{ + any})  \mathcal{B}(\psi(2S) \rightarrow \jpsi\pi^+\pi^-)} = (3.95 \pm 0.32 \mathrm{(stat)} \pm 0.08\mathrm{(sys)})  \times 10^{-2}.
\end{equation}

In the two-lifetime model, the two lifetimes are fixed to expected values for $X(3872)$ originating from the
decays of $B_c$ and from  long-lived $b$-hadrons, respectively, with their relative weight
determined from the fits to the data.
The ratio of the branching fractions $R_B$ is determined from the long-lived component alone:
\begin{equation}
R^{\mathrm{2L}}_{B} =  \frac{\mathcal{B}(B \rightarrow X(3872) \textrm{ + any})  \mathcal{B}(X(3872) \rightarrow
\jpsi\pi^+\pi^-)}{\mathcal{B}(B \rightarrow \psi(2S) \textrm{ + any})  \mathcal{B}(\psi(2S) \rightarrow
\jpsi\pi^+\pi^-)} = (3.57 \pm 0.33 \mathrm{(stat)} \pm 0.11\mathrm{(sys)}) \times 10^{-2} .
\end{equation}

In the two-lifetime model, the fraction of the short-lived non-prompt component in $X(3872)$ production,
for $\pt>10$~GeV, is found to be
\begin{equation}
\frac{\sigma(pp\to B_c \textrm{ + any})\mathcal{B}(B_c\to X(3872)\textrm{ + any})}{\sigma(pp\to
\textrm{non-prompt\ }X(3872)\textrm{ + any})} = (25 \pm 13\mathrm{(stat)} \pm
2\mathrm{(sys)}\pm 5\mathrm{(spin)})\%.
\end{equation}

The invariant mass distributions of the dipion system in $\psi(2S) \to
\jpsi\pi^+\pi^-$ and $X(3872) \rightarrow \jpsi\pi^+\pi^-$ decays are also
measured. The results disfavour a phase-space distribution in both cases,
and point strongly to the dominance of the  $X(3872) \to \jpsi \rho^0$ mode in $X(3872)$ decays.

\section*{Acknowledgements}


We thank CERN for the very successful operation of the LHC, as well as the
support staff from our institutions without whom ATLAS could not be
operated efficiently.

We acknowledge the support of ANPCyT, Argentina; YerPhI, Armenia; ARC, Australia; BMWFW and FWF, Austria; ANAS, Azerbaijan; SSTC, Belarus; CNPq and FAPESP, Brazil; NSERC, NRC and CFI, Canada; CERN; CONICYT, Chile; CAS, MOST and NSFC, China; COLCIENCIAS, Colombia; MSMT CR, MPO CR and VSC CR, Czech Republic; DNRF and DNSRC, Denmark; IN2P3-CNRS, CEA-DSM/IRFU, France; GNSF, Georgia; BMBF, HGF, and MPG, Germany; GSRT, Greece; RGC, Hong Kong SAR, China; ISF, I-CORE and Benoziyo Center, Israel; INFN, Italy; MEXT and JSPS, Japan; CNRST, Morocco; FOM and NWO, Netherlands; RCN, Norway; MNiSW and NCN, Poland; FCT, Portugal; MNE/IFA, Romania; MES of Russia and NRC KI, Russian Federation; JINR; MESTD, Serbia; MSSR, Slovakia; ARRS and MIZ\v{S}, Slovenia; DST/NRF, South Africa; MINECO, Spain; SRC and Wallenberg Foundation, Sweden; SERI, SNSF and Cantons of Bern and Geneva, Switzerland; MOST, Taiwan; TAEK, Turkey; STFC, United Kingdom; DOE and NSF, United States of America. In addition, individual groups and members have received support from BCKDF, the Canada Council, CANARIE, CRC, Compute Canada, FQRNT, and the Ontario Innovation Trust, Canada; EPLANET, ERC, FP7, Horizon 2020 and Marie Sk{\l}odowska-Curie Actions, European Union; Investissements d'Avenir Labex and Idex, ANR, R{\'e}gion Auvergne and Fondation Partager le Savoir, France; DFG and AvH Foundation, Germany; Herakleitos, Thales and Aristeia programmes co-financed by EU-ESF and the Greek NSRF; BSF, GIF and Minerva, Israel; BRF, Norway; Generalitat de Catalunya, Generalitat Valenciana, Spain; the Royal Society and Leverhulme Trust, United Kingdom.

The crucial computing support from all WLCG partners is acknowledged gratefully, in particular from CERN, the ATLAS Tier-1 facilities at TRIUMF (Canada), NDGF (Denmark, Norway, Sweden), CC-IN2P3 (France), KIT/GridKA (Germany), INFN-CNAF (Italy), NL-T1 (Netherlands), PIC (Spain), ASGC (Taiwan), RAL (UK) and BNL (USA), the Tier-2 facilities worldwide and large non-WLCG resource providers. Major contributors of computing resources are listed in Ref.~\cite{ATL-GEN-PUB-2016-002}.

\clearpage
\appendix
\part*{Appendix}
\addcontentsline{toc}{part}{Appendix}

\section{Spin-alignment}
\label{app:spinAllignment}

The acceptance of the $\mu^+\mu^-\pi^+\pi^-$ final state depends on the spin-alignment of the parent state.
Several polarisation hypotheses were considered,
based on the measured quantum numbers of the hidden-charm states
($J^{P}=1^{-}$ for $\psi(2S)$ and $\jpsi$~\cite{PDGBcTau}, $1^+$ for $X(3872)$~\cite{lhcbQN}) and of the dipion system ($0^+$ in $\psi(2S) \to \jpsi\pi^+\pi^-$ decay~\cite{voloshinBES}, $1^-$ in $X(3872) \to \jpsi\pi^+\pi^-$~\cite{lhcbQN}).
In both decays, the dipion system is assumed to be in $S$-wave with respect to the $\jpsi$.

\begin{figure}[H]
\begin{center}
\subfigure[]{\includegraphics[width=0.49\textwidth]{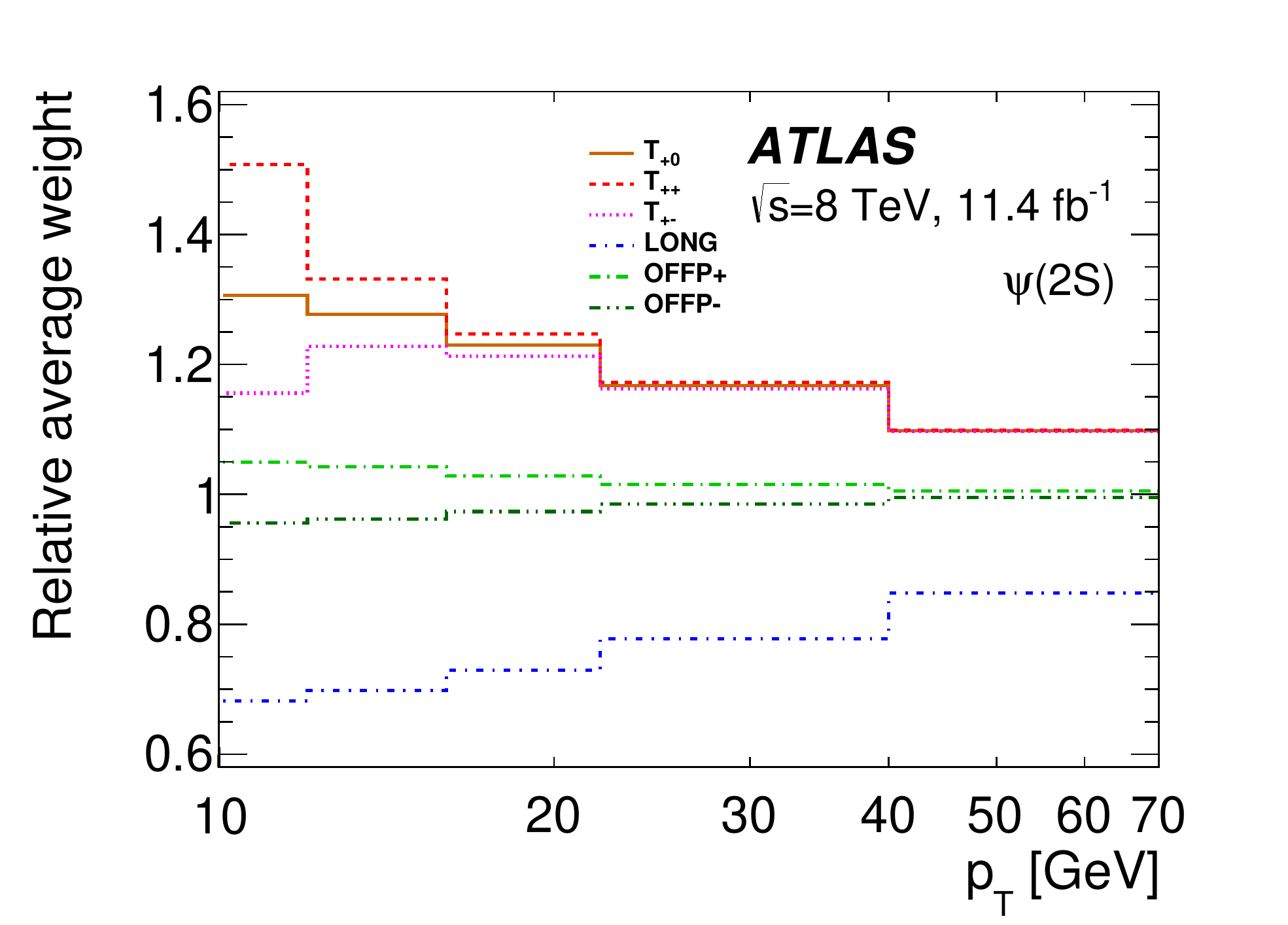}\label{polYieldsPsi2S} }
\subfigure[]{\includegraphics[width=0.49\textwidth]{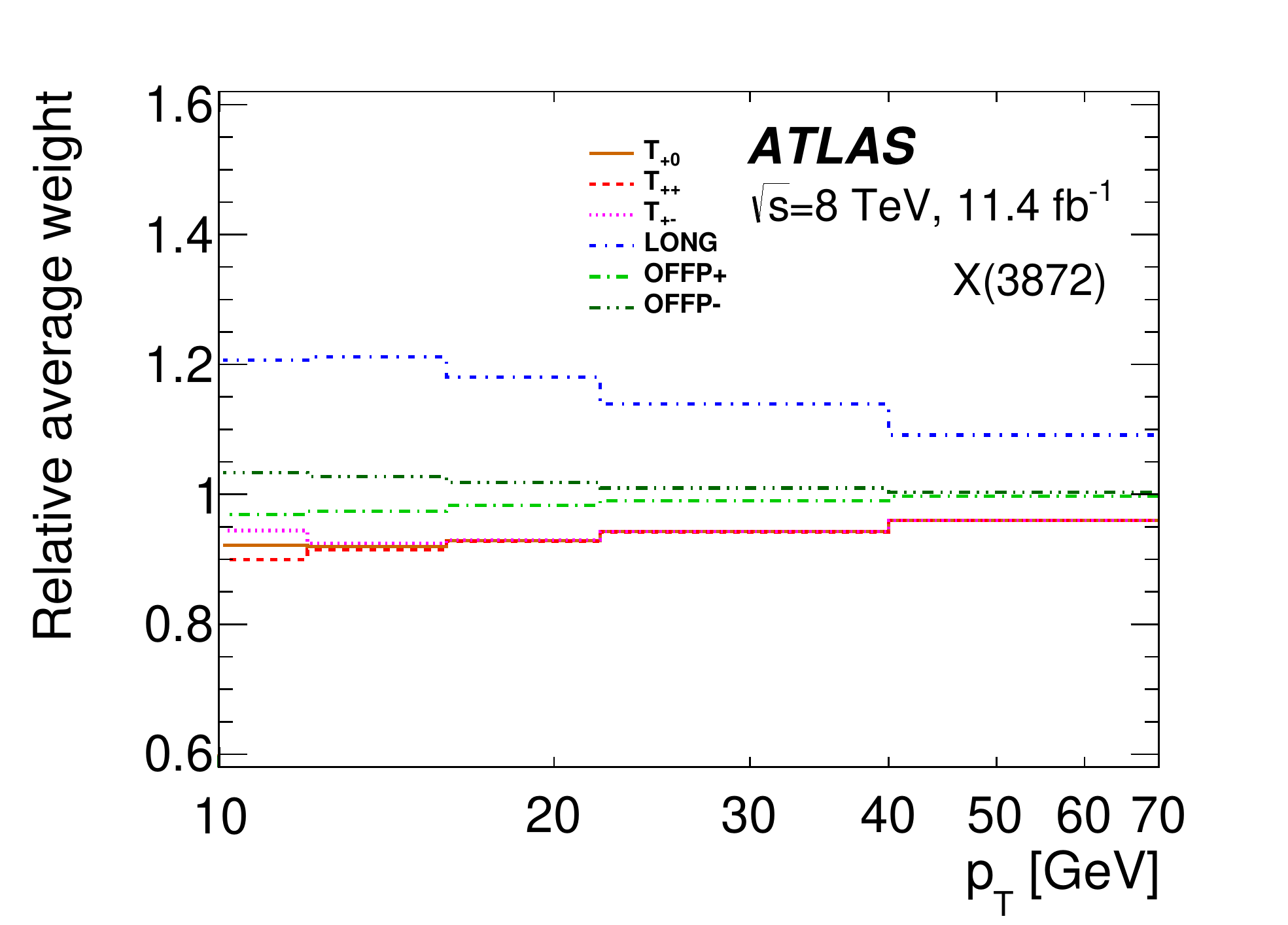}\label{polYieldsX3872} }
\caption{Correction factors for the \subref{polYieldsPsi2S} $\psi(2S)$ and \subref{polYieldsX3872} $X(3872)$ yields for various polarisation hypotheses.}
\label{fig:polYields}
\end{center}
\end{figure}

The spin-alignment scenarios considered in this paper were derived using
the helicity formalism~\cite{Jacob:1959at,Richman:1984gh,Chung:1997jn}, and
are conveniently classified in terms of the various helicity amplitudes of the parent state, $A_m$, with $m=-1, 0, +1$:
\begin{itemize}
\item \textit{Unpolarised - }
an incoherent superposition of $A_{-}=1, A_0=1$ and $A_+=1$, which is labelled UNPOL. This is used as the central hypothesis.
\item \textit{Transversely} polarised with either $A_+ = +1, A_0 = 0, A_-= 0$, or
$A_+ = 0, A_0 = 0, A_-= +1$, which is labelled \textit{T$_{+0}$}.
\item \textit{Transversely} polarised with $A_+ = +1/\sqrt{2}, A_0=0, A_- = +1/\sqrt{2}$, which is labelled \textit{T$_{++}$}.
\item \textit{Transversely} polarised with $A_+ = -1/\sqrt{2}, A_0=0, A_- = +1/\sqrt{2}$, which is labelled \textit{T$_{+-}$}.
\item \textit{Longitudinally} polarised with $A_+ = 0, A_0=+1, A_- = 0$, which is labelled LONG.
\item \textit{Off-Plane Positive - } with $A_+ = -\sqrt{6}/3, A_0=+\sqrt{3}/3, A_- = 0$, which is labelled OFFP$+$.
\item \textit{Off-Plane Negative - } with $A_+ = +\sqrt{6}/3, A_0=+\sqrt{3}/3, A_- = 0$, which is labelled OFFP$-$.
\end{itemize}
Average acceptance weights are calculated for each of these scenarios in each of the analysis $\pt$ bins. The ratios of the average weights for each polarisation scenario to those of the unpolarised case are shown in Figure~\ref{polYieldsPsi2S} for $\psi(2S)$ and Figure~\ref{polYieldsX3872} for $X(3872)$, with the values tabulated in Tables~\ref{tab:polValues_Psi2S} and~\ref{tab:polValues_X3872}, respectively.

No individual production process can lead to an unpolarised vector state, but an unpolarised vector state can be observed due to a superposition of several production subprocesses with different spin-alignments~\cite{polTheoryPaper}. The polarisation of prompt $\psi(2S)$ has been measured by CMS~\cite{cmsPol}
and LHCb~\cite{lhcbPol2014} and it was found that the angular dependence was close to isotropic, justifying the choice of unpolarised production for the central hypothesis. The non-prompt $\psi(2S)$ and $X(3872)$ are unlikely to show significant spin-alignment, since they are produced from a large number of different incoherent exclusive decays of parent $b$-hadrons.

\begin{table}[htbp]
\renewcommand{\arraystretch}{1.05}
\begin{center}
\begin{tabular}{cccccc}
\hline\hline
& \multicolumn{5}{c}{ $\pt$ [GeV]} \\
Polarisation hypothesis & 10--12 & 12--16 & 16--22 & 22--40 & 40--70 \\ \hline
$T_{+0}$ & $ 1.306$ & $ 1.277$ & $1.229$ & $1.168$ & $1.098$ \\
$T_{++}$ & $ 1.508$ & $ 1.331$ & $1.247$ & $1.173$ & $1.099$ \\
$T_{+-}$ & $ 1.156$ & $ 1.228$ & $1.213$ & $1.163$ & $1.097$ \\
LONG & $ 0.682$ & $ 0.698$ & $0.729$ & $0.777$ & $0.848$ \\
OFFP$+$ & $ 1.049$ & $ 1.042$ & $1.028$ & $1.015$ & $1.005$ \\
OFFP$-$ & $ 0.956$ & $ 0.962$ & $0.974$ & $0.985$ & $0.995$ \\
\hline\hline
\end{tabular}
\end{center}
\caption{\label{tab:polValues_Psi2S}Correction factors for various polarisation hypotheses in $p_{\rm T}$ bins for $\psi(2\mathrm{S})$ production. }
\end{table}

\begin{table}[htbp]
\renewcommand{\arraystretch}{1.05}
\begin{center}
\begin{tabular}{cccccc}
\hline\hline
& \multicolumn{5}{c}{ $\pt$ [GeV]} \\
Polarisation hypothesis & 10--12 & 12--16 & 16--22 & 22--40 & 40--70 \\ \hline
$T_{+0}$ & $ 0.921$ & $ 0.920$ & $0.929$ & $0.943$ & $0.960$ \\
$T_{++}$ & $ 0.900$ & $ 0.915$ & $0.928$ & $0.942$ & $0.960$ \\
$T_{+-}$ & $ 0.944$ & $ 0.925$ & $0.930$ & $0.943$ & $0.960$ \\
LONG & $ 1.207$ & $ 1.212$ & $1.181$ & $1.139$ & $1.091$ \\
OFFP$+$ & $ 0.969$ & $ 0.974$ & $0.983$ & $0.990$ & $0.997$ \\
OFFP$-$ & $ 1.033$ & $ 1.027$ & $1.018$ & $1.010$ & $1.003$ \\
\hline\hline
\end{tabular}
\end{center}
\caption{\label{tab:polValues_X3872}Correction factors for various polarisation hypotheses in $p_{\rm T}$ bins for $X(3872)$ production. }
\end{table}

\newpage
\printbibliography

 \newpage 
\begin{flushleft}
{\Large The ATLAS Collaboration}

\bigskip

M.~Aaboud$^\textrm{\scriptsize 136d}$,
G.~Aad$^\textrm{\scriptsize 87}$,
B.~Abbott$^\textrm{\scriptsize 114}$,
J.~Abdallah$^\textrm{\scriptsize 8}$,
O.~Abdinov$^\textrm{\scriptsize 12}$,
B.~Abeloos$^\textrm{\scriptsize 118}$,
R.~Aben$^\textrm{\scriptsize 108}$,
O.S.~AbouZeid$^\textrm{\scriptsize 138}$,
N.L.~Abraham$^\textrm{\scriptsize 152}$,
H.~Abramowicz$^\textrm{\scriptsize 156}$
% ...
\bigskip
\\
% ...
$^{8}$ Department of Physics, The University of Texas at Arlington, Arlington TX, United States of America\\
% ...
$^{12}$ Institute of Physics, Azerbaijan Academy of Sciences, Baku, Azerbaijan\\
% ...
$^{87}$ CPPM, Aix-Marseille Universit{\'e} and CNRS/IN2P3, Marseille, France\\
% ...
$^{108}$ Nikhef National Institute for Subatomic Physics and University of Amsterdam, Amsterdam, Netherlands\\
% ...
$^{114}$ Homer L. Dodge Department of Physics and Astronomy, University of Oklahoma, Norman OK, United States of America\\
% ...
$^{118}$ LAL, Univ. Paris-Sud, CNRS/IN2P3, Universit{\'e} Paris-Saclay, Orsay, France\\
% ...
$^{136}$ $^{(a)}$ Facult{\'e} des Sciences Ain Chock, R{\'e}seau Universitaire de Physique des Hautes Energies - Universit{\'e} Hassan II, Casablanca; $^{(b)}$ Centre National de l'Energie des Sciences Techniques Nucleaires, Rabat; $^{(c)}$ Facult{\'e} des Sciences Semlalia, Universit{\'e} Cadi Ayyad, LPHEA-Marrakech; $^{(d)}$ Facult{\'e} des Sciences, Universit{\'e} Mohamed Premier and LPTPM, Oujda; $^{(e)}$ Facult{\'e} des sciences, Universit{\'e} Mohammed V, Rabat, Morocco\\
% ...
$^{138}$ Santa Cruz Institute for Particle Physics, University of California Santa Cruz, Santa Cruz CA, United States of America\\
% ...
$^{152}$ Department of Physics and Astronomy, University of Sussex, Brighton, United Kingdom\\
% ...
$^{156}$ Raymond and Beverly Sackler School of Physics and Astronomy, Tel Aviv University, Tel Aviv, Israel\\
% ...
$^{a}$ Also at Department of Physics, King's College London, London, United Kingdom\\
$^{b}$ Also at Institute of Physics, Azerbaijan Academy of Sciences, Baku, Azerbaijan\\
$^{c}$ Also at Novosibirsk State University, Novosibirsk, Russia\\
$^{d}$ Also at TRIUMF, Vancouver BC, Canada\\
$^{e}$ Also at Department of Physics {\&} Astronomy, University of Louisville, Louisville, KY, United States of America\\
% ...
\end{flushleft}

 
\end{document}